\documentclass[a4paper,fleqn,usenatbib]{mnras}

\usepackage{newtxtext,newtxmath}
\usepackage[T1]{fontenc}

\usepackage[english]{babel}
\usepackage{graphicx}
\usepackage{fleqn}
\usepackage{amssymb}
\usepackage{amsmath}
\usepackage{booktabs}
\usepackage{hyperref}
\usepackage{comment}
\usepackage{enumitem}
\setlist[enumerate]{
  labelsep=8pt,
  labelindent=0.\parindent,
 itemindent=0pt,
  leftmargin=*,
}

\usepackage{color, colortbl}

\newcommand*{\mycdot}{\kern-.2em\cdot\kern-.2em}
\renewcommand{\S}{Section}
\newcommand{\F}{Fig.}
\newcommand{\eq}{equation}
\newcommand{\Eq}{Equation}

\newcommand{\ve}[1]{\boldsymbol{#1}}
\newcommand{\unit}[1]{\hat{\boldsymbol{#1}}}

\newcommand{\msun}{\mathrm{M}_\odot}

\newcommand{\au}{\,\textsc{au}}

\renewcommand{\j}{\jmath}

\newcommand{\epssa}[1]{\epsilon_{\mathrm{SA},#1}}

\newcommand{\epsoct}[1]{\epsilon_{\mathrm{oct},#1}}
\newcommand{\epshex}[1]{\epsilon_{\mathrm{hex},#1}}
\newcommand{\epshexcross}[1]{\epsilon_{\mathrm{hex,cross},#1}}

\newcommand{\eper}{E}

\newcommand{\qper}{Q}

\pubyear{2020}

\usepackage{listings}

\usepackage{color}

\definecolor{dkgreen}{rgb}{0,0.6,0}
\definecolor{gray}{rgb}{0.5,0.5,0.5}
\definecolor{mauve}{rgb}{0.58,0,0.82}

\allowdisplaybreaks

\begin{document}
\onecolumn
\title[Binary-binary encounters]{Binary-binary scattering in the secular limit}

\author[Hamers \& Samsing]{Adrian S. Hamers$^{1}$\thanks{E-mail: hamers@mpa-garching.mpg.de} and Johan Samsing$^{2}$\thanks{E-mail: jsamsing@gmail.com} \\
$^{1}$Max-Planck-Institut f\"{u}r Astrophysik, Karl-Schwarzschild-Str. 1, 85741 Garching, Germany \\
$^{2}$Niels Bohr International Academy, The Niels Bohr Institute, Blegdamsvej 17, DK-2100, Copenhagen , Denmark}
\date{Accepted 2020 March 9. Received 2020 March 6; in original form 2020 February 12}

\label{firstpage}
\pagerange{\pageref{firstpage}--\pageref{lastpage}}
\maketitle

\begin{abstract}  
Binary-binary interactions are important in a number of astrophysical contexts including dense stellar systems such as globular clusters. Although less frequent than binary-single encounters, binary-binary interactions lead to a much richer range of possibilities such as the formation of stable triple systems. Here, we focus on the regime of distant binary-binary encounters, i.e., two binaries approaching each other on an unbound orbit with a periapsis distance $\qper$ much larger than the internal binary separations. This `secular' regime gives rise to changes in the orbital eccentricities and orientations, which we study using analytic considerations and numerical integrations. We show that `direct' interactions between the three orbits only occur starting at a high expansion order of the Hamiltonian (hexadecupole order), and that the backreaction of the outer orbit on the inner two orbits at lower expansion orders is weak. Therefore, to good approximation, one can obtain the changes of each orbit by using previously-known analytic results for binary-single interactions, and replacing the mass of the third body with the total mass of the companion binary. Nevertheless, we find some dependence of the `binarity' of the companion binary, and derive explicit analytic expressions for the secular changes that are consistent with numerical integrations. In particular, the eccentricity and inclination changes of orbit 1 due to orbit 2 scale as $\epssa{1} (a_2/\qper)^2 [m_3 m_4/(m_3+m_4)^2]$, where $\epssa{1}$ is the approximate quadrupole-order change, and $a_2$ and $(m_3,m_4)$ are the companion binary orbital semimajor axis and component masses, respectively. Our results are implemented in several \textsc{Python} scripts that are freely available. 
\end{abstract}


\section{Introduction}
\label{sect:introduction}
Dense stellar systems such as open and globular clusters are host to a wide range of dynamical interactions involving bound objects such as binaries-single scattering, as well as scattering involving higher-order systems, e.g., binary-binary scattering. Since such interactions are believed to lead to mergers of black holes (BHs) and neutron stars (NSs) (e.g., \citealt{1993Natur.364..423S,2000ApJ...528L..17P,2006ApJ...637..937O,2014MNRAS.441.3703Z,2015PhRvL.115e1101R,
2016PhRvD..93h4029R,2016MNRAS.463.2443K,2016MNRAS.459.3432M,
2017ApJ...840L..14S,2018ApJ...853..140S,2018ApJ...855..124S,2018PhRvD..97j3014S,2018PhRvL.120o1101R,2019PhRvD.100d3010S}, interest in them has recently surged with the direct detection of gravitational waves (GWs) from merging black holes (BHs) and neutron stars (NSs; e.g., \citealt{2016PhRvL.116x1103A,2016PhRvL.116f1102A,2017PhRvL.118v1101A,2017ApJ...851L..35A,2017PhRvL.119n1101A,2017ApJ...848L..12A}).

The topic of binary-single scattering has received a great deal of attention in the past decades (e.g., \citealt{1983ApJ...268..319H,1983ApJ...268..342H,1993ApJS...85..347H,1993ApJ...403..256H,1993ApJ...403..271G,1993ApJ...415..631S,1993ApJ...411..285D,1996ApJ...467..348M,1996ApJ...467..359H,2012PhRvD..85l3005K,2018arXiv180208654S}). Binary-binary encounters have been studied as well, although perhaps with less intensity given its greater complexity. Nevertheless, in star clusters with binary fractions $\gtrsim 10\%$, binary-binary interactions dominate over binary-single interactions \citep{1993ApJ...415..631S,2011MNRAS.410.2370L}. Furthermore, even if the overall binary fraction of a stellar cluster is low, the binary fraction in the core can be much higher \citep{1989AJ.....98..217L,1992ApJ...389..527H,1994ApJ...427..793M}. Binary-binary scattering can also occur in other astrophysical contexts, such as binaries passing by planetary systems in the field (e.g., \citealt{2015MNRAS.448..344L}). 

Studies of binary-binary scattering to date (e.g., \citealt{1983MNRAS.203.1107M,1983AJ.....88.1420H,1984MNRAS.207..115M,1984MNRAS.208...75M,1986JCoPh..64..195A,1989AJ.....98..217L,1995ApJ...438L..33R,1996MNRAS.281..830B,2012MNRAS.425.2369L,2015MNRAS.450.1724L,2016MNRAS.456.4219A,2017MNRAS.471.1830L,2017MNRAS.467.4447R,2018MNRAS.480.3062L,2019ApJ...871...91Z}) have mostly focussed on numerical investigations of ``strong'' scattering, i.e., when the two binaries approach each other sufficiently closely that at least their binding energies change appreciably, and, more generally, leading to complex interactions such as the breakup of binaries, exchange interactions, and the formation of (stable or unstable) triples. 

However, more distant encounters are more common than the close encounters that give rise to ``strong'' interactions. In these more distant encounters with periapsis distances $\qper\gg a_i$, where $a_i$ ($i\in\{1,2\}$) are the semimajor axes of the two bound binaries, energy changes are exponentially suppressed \citep{1975MNRAS.173..729H}, whereas angular-momentum changes can still occur. These more distant encounters can be characterised as `secular', i.e., the orbital motion of the components in the bound systems is much faster than the orbital motion of the wider, unbound orbit. These secular encounters have been studied by a number of authors in the context of binary-single encounters (e.g., \citealt{1996MNRAS.282.1064H,2009ApJ...697..458S,2018MNRAS.476.4139H,2019ApJ...872..165G,2019MNRAS.487.5630H,2019MNRAS.488.5192H}). Secular binary-single encounters can have important implications for the properties of binary BH mergers in globular clusters \citep{2019PhRvD.100d3010S}. However, to our knowledge, secular effects in binary-binary encounters have not been addressed before.

In this paper, we consider the dynamical evolution of two binaries approaching each other on a parabolic or hyperbolic orbit with a periapsis distance larger than the binaries' internal separations. In \S~\ref{sect:an},  we derive expressions for the secular changes in the two binaries based on the expanded and partially-averaged Hamiltonian of the system. In \S~\ref{sect:num}, we carry out numerical simulations (direct-integration four-body simulations, as well as semianalytic integrations based on the partially-averaged Hamiltonian) and use these to test our analytic expressions. We discuss our results in \S~\ref{sect:discussion}, and conclude in \S~\ref{sect:conclusions}.

\section{Analytic considerations}
\label{sect:an}
\subsection{Setup}
\label{sect:an:setup}

\begin{figure}
\center
\includegraphics[scale = 0.65, trim = 0mm 10mm 0mm 10mm]{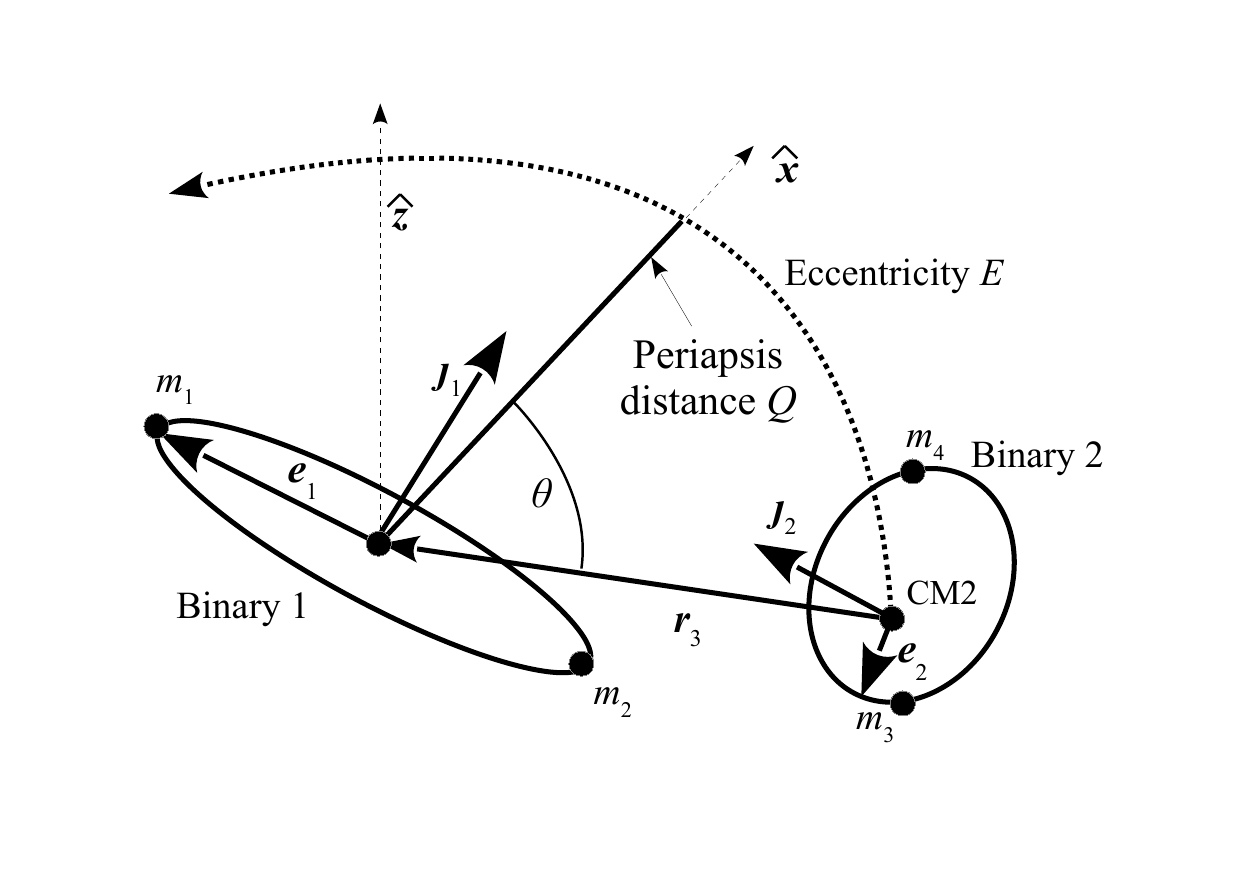}
\caption{Sketch of the configuration. Two bound binaries (labeled `1' and `2') approach each other on a parabolic or hyperbolic orbit with eccentricity $\eper\geq1$ and periapsis distance $Q>a_i$ ($i\in\{1,2\}$). }
\label{fig:sketch}
\end{figure}

We consider two bound binary systems (their orbits indicated with `1' and `2') that approach each other on an unbound orbit (the latter is referred to as orbit `3', or the `outer' orbit). See \F~\ref{fig:sketch} for a sketch. Let the masses of the components in orbit 1 be $m_1$ and $m_2$, respectively, and $m_3$ and $m_4$ for orbit 2. For convenience, we introduce the quantities $M_1 \equiv m_1+m_2$, $M_2 \equiv m_3+m_4$, and $M_3 \equiv M\equiv M_1+M_2$. The (initial) semimajor axes and eccentricities of all orbits are denoted with $a_i$ and $e_i$, where it should noted that $a_3 < 0$, and $e_3 \geq 1$. Also, to further distinguish between the bound and unbound orbits, we introduce the notation $\eper \equiv e_3\geq 1$ for the (initial) outer orbit eccentricity, and $\qper\equiv a_3(1-e_3)>0$ for the outer orbit periapsis distance. Note that in defining $\qper$ and $\eper$, we neglect the extended nature of the bound orbits (i.e., approximate the latter as point masses). Evidently, since we are dealing with two binaries instead of two point particles, the latter approximation breaks down as $\qper\rightarrow 0$. 

Let the relative separation between the two bodies in orbit 1 be denoted with $\ve{r}_1$; similarly, the relative separation vector between bodies 3 and 4 in orbit 2 is $\ve{r}_2$. The outer orbit has a separation vector $\ve{r}_3$ between the two centers of mass of orbits 1 and 2. The instantaneous eccentricity or Laplace-Runge-Lenz vector $\ve{e}_i$ is given by $\ve{e}_i = [1/(GM_i)] \, \dot{\ve{r}}_i\times (\ve{r}_i\times \dot{\ve{r}}_i) - \unit{r}_i$, where dots denote derivatives with respect to time, and hats denote unit vectors. For orbits 1 and 2, the normalized angular-momentum vectors are $\ve{\j}_i=\ve{r}_i \times \dot{\ve{r}}_i$, with magnitudes $\j_i=\sqrt{1-e_i^2}$. 

Without loss of generality, we assume that the outer orbit is initially oriented with its angular-momentum vector along the $z$-axis, and the periapsis pointing along the $x$-axis. In this case, and neglecting the backreaction of the outer orbit due to the quadrupole moment of the inner orbits (see \S~\ref{sect:an:an:out} below), the outer orbit is described according to
\begin{subequations}
\begin{align}
\ve{r}_3 &= \frac{\qper (1 + \eper)}{1 + \eper \cos \theta} \left [ \cos \theta \, \unit{x} + \sin \theta \, \unit{y} \right ]; \\
\dot{\ve{r}}_3 &= \sqrt{\frac{GM}{\qper(1+\eper)}} \left [ - \sin \theta \, \unit{x} + \left(\eper+\cos\theta \right ) \, \unit{y} \right ].
\end{align}
\end{subequations}
The outer orbit true anomaly $\theta$ is related to the physical time $t$ according to
\begin{align}
\label{eq:thetat}
\mathrm{d} t = \frac{1}{n_3} \left (\eper^2-1 \right ) \frac{1}{(1+\eper \cos \theta )^2} \mathrm{d} \theta,
\end{align}
where $n_3 \equiv \sqrt{GM/|a_3|^3}$ is the hyperbolic mean motion, and $|a_3| = \qper/(\eper-1)$. The true anomaly $\theta$ ranges between $-L$ and $L$ corresponding to $-\infty< t < \infty$, where
\begin{align}
L \equiv \arccos \left ( -\frac{1}{\eper} \right ).
\end{align}

\subsection{Hamiltonian}
\label{sect:an:ham}
\subsubsection{Expansion}
\label{sect:an:ham:ex}
In the limit that the two binaries approach each other with a periapsis distance $\qper\gg r_1,r_2$, it is appropriate to expand the Hamiltonian of the four-body system in terms of the small ratios $r_1/r_3\ll1$, and $r_2/r_3\ll1$. The resulting `binary-binary' Hamiltonian is (\citealt{2015MNRAS.449.4221H}, see also \citealt{2016MNRAS.459.2827H})
\begin{align}
\nonumber H_\mathrm{bb} &= H_\mathrm{kep}(m_1,m_2,\ve{r}_1,\dot{\ve{r}}_1) + H_\mathrm{kep}(m_3,m_4,\ve{r}_2,\dot{\ve{r}}_2) + H_\mathrm{kep}(m_1+m_2,m_3+m_4,\ve{r}_3,\dot{\ve{r}}_3) \\
\nonumber &\quad + H_\mathrm{quad} (m_1,m_2,m_3+m_4,\ve{r}_1,\ve{r}_3) + H_\mathrm{quad} (m_3,m_4,m_1+m_2,\ve{r}_2,\ve{r}_3) \\
\nonumber &\quad + H_\mathrm{oct} (m_1,m_2,m_3+m_4,\ve{r}_1,\ve{r}_3) + H_\mathrm{oct} (m_4,m_3,m_1+m_2,\ve{r}_2,\ve{r}_3) \\
\nonumber &\quad + H_\mathrm{hd}(m_1,m_2,m_3+m_4,\ve{r}_1,\ve{r}_3) + H_\mathrm{hd}(m_4,m_3,m_1+m_2,\ve{r}_2,\ve{r}_3)  + H_\mathrm{hd,\,cross,\,bb}(m_1,m_2,m_3,m_4,\ve{r}_1,\ve{r}_2,\ve{r}_3) \\
&\quad + \mathcal{O} \left [ \frac{1}{r_3} \left ( \frac{r_1}{r_2}\right)^i \left ( \frac{r_2}{r_3}\right)^j \left (\frac{r_1}{r_3}\right)^k \right ].
\label{eq:H_noav}
\end{align}
Here, $i+j+k\geq 5$. The various `universal' functions in \eq~(\ref{eq:H_noav}) are given by
\begin{subequations}
\label{eq:app:Hfuncts}
\begin{align}
\label{eq:app:H_bin_non_av}
&H_\mathrm{kep}(m,m',\boldsymbol{r},\dot{\boldsymbol{r}}) = \frac{1}{2} \frac{mm'}{m+m'} \dot{\boldsymbol{r}}^2 - \frac{Gmm'}{r}; \\
\label{eq:app:H_quad_non_av}
&H_\mathrm{quad}(m,m',m'',\boldsymbol{r},\boldsymbol{r}') = -\frac{Gmm'm''}{m+m'} \frac{1}{r'} \left (\frac{r}{r'} \right )^2 \frac{1}{2} \left [ 3 \left (\hat{\boldsymbol{r}}\cdot \hat{\boldsymbol{r}}' \right )^2 - 1 \right ]; \\
\label{eq:app:H_oct_non_av}
&H_\mathrm{oct}(m,m',m'',\boldsymbol{r},\boldsymbol{r}') = -\frac{Gmm'm''(m-m')}{(m+m')^2} \frac{1}{r'} \left (\frac{r}{r'} \right )^3 \frac{1}{2} \left [ 5 \left (\hat{\boldsymbol{r}}\cdot \hat{\boldsymbol{r}}' \right )^3 - 3 \left (\hat{\boldsymbol{r}}\cdot \hat{\boldsymbol{r}}' \right ) \right ]; \\
\label{eq:app:H_hd_non_av}
&H_\mathrm{hd}(m,m',m'',\boldsymbol{r},\boldsymbol{r}') = -\frac{Gmm'm''(m^2-mm'+m'^2)}{(m+m')^3} \frac{1}{r'} \left (\frac{r}{r'} \right )^4 \frac{1}{8} \left [ 35 \left (\hat{\boldsymbol{r}}\cdot \hat{\boldsymbol{r}}' \right )^4 - 30  \left (\hat{\boldsymbol{r}}\cdot \hat{\boldsymbol{r}}' \right )^2 + 3 \right ]; \\
\nonumber &H_\mathrm{hd,\,cross,\,bb}(m_1,m_2,m_3,m_4,\ve{r}_1,\ve{r}_2,\ve{r}_3) = - \frac{Gm_1m_2m_3m_4}{(m_1+m_2)(m_3+m_4)} \frac{1}{r_3} \left ( \frac{r_1}{r_3} \right )^2 \left ( \frac{r_2}{r_3} \right )^2 \\
&\quad \times \frac{3}{4} \left [ 1 - 5 \left( \unit{r}_1 \cdot \unit{r}_3 \right )^2 - 5 \left( \unit{r}_2 \cdot \unit{r}_3 \right )^2 + 35 \left( \unit{r}_1 \cdot \unit{r}_3 \right )^2 \left( \unit{r}_2 \cdot \unit{r}_3 \right )^2 + 2 \left( \unit{r}_1 \cdot \unit{r}_2 \right )^2 -20 \left( \unit{r}_1 \cdot \unit{r}_2 \right ) \left( \unit{r}_1 \cdot \unit{r}_3 \right ) \left( \unit{r}_2 \cdot \unit{r}_3 \right )\right ].
\label{eq:app:H_hd_cross_bb}
\end{align}
\end{subequations}

The first three terms in \Eq~(\ref{eq:H_noav}) are the Keplerian terms; in the limit that the orbits are described by Keplerian orbits, these terms individually reduce to the constant terms $-G(m + m')/(2a_i)$, where $i$ refers to the corresponding orbit and the masses should be replaced appropriately for each orbit. The other terms in \eq~(\ref{eq:H_noav}) give rise to changes to the Keplerian orbits. Note that, as expected, $H_\mathrm{bb}$ is symmetric with respect to binaries 1 and 2, i.e., it is invariant under the interchange of parameters $m_1 \leftrightarrow m_3$, $m_2 \leftrightarrow m_4$, and $\ve{r}_1 \leftrightarrow \ve{r}_2$. 

It is immediately clear that, to the lowest expansion orders, the quadrupole and octupole orders, $\propto (r/r')^2$ and $\propto (r/r')^3$, respectively, the expanded Hamiltonian is fully described in terms of pairwise interactions only: the interaction between orbit 1 and its outer orbit, and between orbit 2 and its outer orbit. This implies that, up to and including octupole order, any effect of the `binarity' of the companion orbit can only arise from an effect of the companion binarity on the outer orbit, i.e., on $\ve{r}_3$. This is discussed further below analytically (\S~\ref{sect:an:an:out}), as well as numerically (\S~\ref{sect:num:out}).

Only at the `hexadecupole' order, $\propto (r/r')^4$, does there appear a term that involves the properties of three orbits simultaneously, described by $H_\mathrm{hd,\,cross,\,bb}$. We remark that the latter term contains the factors $(r_1/r_3)^2$ and $(r_2/r_3)^2$ which, individually considered, might suggest that $H_\mathrm{hd,\,cross,\,bb}$ should be counted as a quadrupole-order term. However, since we assume that both $r_1/r_3 \ll1$ {\it and} $r_2/r_3\ll 1$, the term $H_\mathrm{hd,\,cross,\,bb}$ is effectively of fourth order; therefore, we consider it a hexadecupole-order term. 

\subsubsection{Partial orbit averaging}
\label{sect:an:ham:oa}
In the `secular' approximation, one averages the expanded Hamiltonian, \eq~(\ref{eq:H_noav}), over some or all orbits. Here, we choose to average over the `inner' orbits, i.e., orbits 1 and 2. This approximation is generally expected to be a good one if $\mathcal{R}_i \ll 1$, where $i\in\{1,2\}$ refers to orbits 1 and 2, and
\begin{align}
\label{eq:Ri}
\mathcal{R}_i = \left [ \left ( 1 + \frac{M_{3-i}}{M_i} \right ) \left (\frac{a_i}{\qper} \right )^3 \left ( 1 + \eper \right ) \right ]^{1/2}.
\end{align}
If $\mathcal{R}_i\ll1$ for both $i=1$ and $i=2$, the mean motions of both bound orbits are much faster than the angular speed of the para/hyperbolic orbit at periapsis (this consideration is analogous to the binary-single case; see, e.g., \eq~1 of \citealt{2019MNRAS.487.5630H}). 

The result of the `inner' averaging, written explicitly to the same order as in \eq~(\ref{eq:H_noav}), is (see, e.g., \citealt{2018MNRAS.476.4139H} for a general derivation of the pairwise averaged expressions to any expansion order; the term $\overline{H}_\mathrm{hex,\,cross,\,bb}$ is derived new here)
\begin{align}
\nonumber \overline{H}_\mathrm{bb} &= \overline{H}_\mathrm{kep}(m_1,m_2,a_1) +\overline{H}_\mathrm{kep}(m_3,m_4,a_2) + \overline{H}_\mathrm{kep}(m_1+m_2,m_3+m_4,a_3) \\
\nonumber &\quad+  \overline{H}_\mathrm{quad} (m_1,m_2,m_3+m_4,a_1,\ve{e}_1,\ve{\j}_1,\ve{r}_3) + \overline{H}_\mathrm{quad} (m_3,m_4,m_1+m_2,a_2,\ve{e}_2,\ve{\j}_2,\ve{r}_3) \\
\nonumber &\quad + \overline{H}_\mathrm{oct} (m_1,m_2,m_3+m_4,a_1,\ve{e}_1,\ve{\j}_1,\ve{r}_3) + \overline{H}_\mathrm{oct} (m_3,m_4,m_1+m_2,a_2,\ve{e}_2,\ve{\j}_2,\ve{r}_3) \\
\nonumber &\quad + \overline{H}_\mathrm{hex} (m_1,m_2,m_3+m_4,a_1,\ve{e}_1,\ve{\j}_1,\ve{r}_3) + \overline{H}_\mathrm{hex} (m_3,m_4,m_1+m_2,a_2,\ve{e}_2,\ve{\j}_2,\ve{r}_3) + \overline{H}_\mathrm{hex,\,cross,\,bb} (m_1,m_2,m_3,m_4,a_1,\ve{e}_1,\ve{\j}_1,a_2,\ve{e}_2,\ve{\j}_2,\ve{r}_3) \\
&\quad + \mathcal{O} \left [ \frac{1}{r_3} \left ( \frac{a_1}{a_2}\right)^i \left ( \frac{a_2}{r_3}\right)^j \left (\frac{a_1}{r_3}\right)^k \right ].
\label{eq:H_av}
\end{align}
Here, we defined
\begin{subequations}
\begin{align}
&\overline{H}_\mathrm{kep}(m,m',a) \equiv - \frac{Gmm'}{2a}; \\
& \overline{H}_\mathrm{quad} (m,m',m'',a,\ve{e},\ve{\j},\ve{r}) \equiv - \frac{G m m' m''}{m+m'} \frac{1}{r} \left ( \frac{a}{r} \right )^2 \frac{1}{4} \left [ 1 - 6 e^2 + 15 \left ( \ve{e} \cdot \unit{r} \right )^2 - 3 \left ( \ve{\j} \cdot \unit{r} \right )^2 \right ]; \\
& \overline{H}_\mathrm{oct} (m,m',m'',a,\ve{e},\ve{\j},\ve{r}) \equiv - \frac{G m m' m''}{m+m'} \frac{|m-m'|}{m+m'} \frac{1}{r} \left ( \frac{a}{r} \right )^3 \frac{5}{16} \left ( \ve{e} \cdot \unit{r} \right ) \left [ 3\left(1-8e^2 \right ) + 35 \left(\ve{e}\cdot \unit{r} \right )^2 - 15 \left (\ve{\j} \cdot \unit{r} \right )^2 \right ]; \\
&\nonumber \overline{H}_\mathrm{hd} (m,m',m'',a,\ve{e},\ve{\j},\ve{r}) \equiv - \frac{G m m' m''}{m+m'} \frac{m^2-m m'+m'^2}{m+m'} \frac{1}{r} \left ( \frac{a}{r} \right )^4 \frac{3}{64} \left [ 3-20e^2+80e^4 + 735 \left(\ve{e} \cdot \unit{r}\right)^4 + 35 \left ( \ve{\j} \cdot \unit{r} \right )^4  - 10 \left(3-10e^2 \right ) \left (\ve{\j} \cdot \unit{r} \right )^2 \right. \\
&\qquad \left. - 70 \left(\ve{e} \cdot \unit{r} \right )^2 \left\{ 7 \left (\ve{\j} \cdot \unit{r} \right )^2 + 10e^2 - 1 \right \} \right ]; \\
&\nonumber \overline{H}_\mathrm{hex,\,cross,\,bb} (m_1,m_2,m_3,m_4,a_1,\ve{e}_1,\ve{\j}_1,a_2,\ve{e}_2,\ve{\j}_2,\ve{r}) \equiv - \frac{G m_1m_2m_3m_4}{(m_1+m_2)(m_3+m_4)} \frac{1}{r} \left ( \frac{a_1}{r} \right )^2 \left ( \frac{a_2}{r} \right )^2 \frac{3}{16} \left [  1-6e_1^2 - 6e_2^2 + 36 e_1^2 e_2^2 + 50 \left( \ve{e}_1 \cdot \ve{e}_2 \right )^2 \right. \\
&\nonumber \qquad \left. -10 \left ( \ve{e}_1 \cdot \ve{\j}_2 \right )^2 - 10 \left ( \ve{\j}_1 \cdot \ve{e}_2 \right )^2 + 2 \left ( \ve{\j}_1 \cdot \ve{\j}_2 \right )^2 + 25 \left ( \ve{e}_1 \cdot \unit{r} \right )^2 + 25 \left ( \ve{e}_2 \cdot \unit{r} \right )^2 +5 \left(6e_2^2-1 \right ) \left ( \ve{\j}_1 \cdot \unit{r} \right)^2 + 5 \left (6e_1^2-1 \right ) \left (\ve{\j}_2 \cdot \unit{r} \right )^2 \right. \\
&\nonumber \qquad \left. -150 e_2^2 \left ( \ve{e}_1 \cdot \unit{r} \right )^2 - 150 e_1^2 \left ( \ve{e}_2 \cdot \unit{r} \right )^2 - 500 \left ( \ve{e}_1 \cdot \ve{e}_2 \right ) \left (\ve{e}_1 \cdot \unit{r} \right ) \left ( \ve{e}_2 \cdot \unit{r} \right ) + 100 \left ( \ve{\j}_1 \cdot \ve{e}_2 \right ) \left ( \ve{\j}_1 \cdot \unit{r} \right ) \left ( \ve{e}_2 \cdot \unit{r} \right ) + 100 \left ( \ve{e}_1 \cdot \ve{\j}_2 \right ) \left ( \ve{e}_1 \cdot \unit{r} \right ) \left ( \ve{\j}_2 \cdot \unit{r} \right ) \right. \\
&\nonumber  \qquad \left.  - 20 \left ( \ve{\j}_1 \cdot \ve{\j}_2 \right ) \left (\ve{\j}_1 \cdot \unit{r} \right ) \left ( \ve{\j}_2 \cdot \unit{r} \right ) -175 \left ( \ve{\j}_1 \cdot \unit{r} \right )^2 \left ( \ve{e}_2 \cdot \unit{r} \right )^2 - 175 \left (\ve{e}_1 \cdot \unit{r} \right )^2 \left ( \ve{\j}_2 \cdot \unit{r} \right )^2 +35 \left ( \ve{\j}_1 \cdot \unit{r} \right )^2 \left (\ve{\j}_2 \cdot \unit{r} \right )^2 + 875 \left ( \ve{e}_1 \cdot \unit{r} \right )^2 \left ( \ve{e}_2 \cdot \unit{r} \right )^2 \right ].
\end{align}
\end{subequations}
As should be, the inner-averaged Hamiltonian, \eq~(\ref{eq:H_av}), is still symmetric with respect to orbits 1 and 2. 

\subsection{Inner-averaged equations of motion}
\label{sect:an:eom}
Hamilton's equations applied to the inner-averaged Hamiltonian \eq~(\ref{sect:an:ham:oa}) imply the following set of equations of motion for the eccentricity $\ve{e}_i$ and angular momentum $\ve{\j}_i$ vectors of orbits 1 and 2, as well as the equation of motion for the outer orbital separation $\ve{r}_3$:
\begin{subequations}
\label{eq:eom}
\begin{align}
\nonumber \frac{\mathrm{d}\ve{e}_i}{\mathrm{d} \theta} &= \epssa{i} (1+\eper \cos \theta) \, \ve{f}_{\dot{\ve{e}},\mathrm{quad}}(\ve{e}_i,\ve{\j}_i,\unit{r}_3) + \epssa{i} \epsoct{i} (1+\eper \cos \theta)^2 \ve{f}_{\dot{\ve{e}},\mathrm{oct}}(\ve{e}_i,\ve{\j}_i,\unit{r}_3) + \epssa{i} \epshex{i} (1+\eper \cos \theta)^3 \ve{f}_{\dot{\ve{e}},\mathrm{hex}} (\ve{e}_i,\ve{\j}_i,\unit{r}_3) \\
\label{eq:eom:ed} &\quad + \epssa{i} \epshexcross{i} (1+\eper \cos \theta)^3 \ve{f}_{\dot{\ve{e}},\mathrm{hex,cross}}(\ve{e}_i,\ve{\j}_i,\ve{e}_{3-i},\ve{\j}_{3-i},\unit{r}_3) + \dots; \\
\nonumber \frac{\mathrm{d}\ve{\j}_i}{\mathrm{d} \theta} &= \epssa{i} (1+\eper \cos \theta) \, \ve{f}_{\dot{\ve{\j}},\mathrm{quad}}(\ve{e}_i,\ve{\j}_i,\unit{r}_3) + \epssa{i} \epsoct{i} (1+\eper \cos \theta)^2 \ve{f}_{\dot{\ve{\j}},\mathrm{oct}}(\ve{e}_i,\ve{\j}_i,\unit{r}_3) + \epssa{i} \epshex{i} (1+\eper \cos \theta)^3 \ve{f}_{\dot{\ve{\j}},\mathrm{hex}} (\ve{e}_i,\ve{\j}_i,\unit{r}_3) \\
\label{eq:eom:jd} &\quad + \epssa{i} \epshexcross{i} (1+\eper \cos \theta)^3 \ve{f}_{\dot{\ve{\j}},\mathrm{hex,cross}}(\ve{e}_i,\ve{\j}_i,\ve{e}_{3-i},\ve{\j}_{3-i},\unit{r}_3) + \dots; \\
\label{eq:eom:rdd} \frac{\mathrm{d}^2 \ve{r}_3}{\mathrm{d} t^2} &= -\frac{GM}{r_3^3} \ve{r}_3 + \frac{GM}{r_3^2} \ve{f}_{\ddot{\ve{r}}_3} (m_1,m_2,a_1,\ve{e}_1,\ve{\j}_1,\ve{r}_3) + \frac{GM}{r_3^2} \ve{f}_{\ddot{\ve{r}}_3} (m_3,m_4,a_2,\ve{e}_2,\ve{\j}_2,\ve{r}_3) + \dots.
\end{align}
\end{subequations}
Here, `$\dots$' denotes higher-order expansion terms, and the auxiliary functions are defined according to
\begin{subequations}
\begin{align}
&\ve{f}_{\dot{\ve{e}},\mathrm{quad}}(\ve{e},\ve{\j},\unit{r}_3) \equiv \left [-3 \left(\ve{\j}\times \ve{e}\right) - \frac{3}{2} \left (\ve{\j} \cdot \unit{r}_3 \right) \left(\ve{e} \times \unit{r}_3 
\right )+ \frac{15}{2} \left( \ve{e}\cdot \unit{r}_3 \right ) \left ( \ve{\j} \times \unit{r}_3 \right )\right ]; \\
&\ve{f}_{\dot{\ve{\j}},\mathrm{quad}}(\ve{e},\ve{\j},\unit{r}_3) \equiv \left [-\frac{3}{2}  \left ( \ve{\j} \cdot \unit{r}_3 \right ) \left( \ve{\j} \times \unit{r}_3 \right ) + \frac{15}{2} \left ( \ve{e} \cdot \unit{r}_3 \right ) \left ( \ve{e} \times \unit{r}_3 \right ) \right ]; \\
& \ve{f}_{\dot{\ve{e}},\mathrm{oct}}(\ve{e},\ve{\j},\unit{r}_3) \equiv \frac{15}{16} {\footnotesize \Biggl [ - 16 \left(\ve{e} \cdot \unit{r}_3 \right ) \left(\ve{\j}\times \ve{e} \right ) + \left(1-8e^2 \right) \left(\ve{\j} \times \unit{r}_3 \right ) - 10 \left ( \ve{e} \cdot \unit{r}_3 \right ) \left ( \ve{\j} \cdot \unit{r}_3 \right ) \left ( \ve{e} \times \unit{r}_3 \right ) - 5 \left(\ve{\j}\cdot \unit{r}_3 \right )^2 \left ( \ve{\j}\times \unit{r}_3 \right ) + 35 \left ( \ve{e} \cdot \unit{r}_3 \right )^2 \left ( \ve{\j} \times \unit{r}_3 \right ) \Biggl ];} \\
& \ve{f}_{\dot{\ve{e}},\mathrm{oct}}(\ve{\j},\ve{\j},\unit{r}_3) \equiv  \frac{15}{16} \Biggl [ \left(1-8e^2 \right) \left(\ve{e} \times \unit{r}_3 \right ) - 10 \left ( \ve{e} \cdot \unit{r}_3 \right ) \left ( \ve{\j} \cdot \unit{r}_3 \right ) \left ( \ve{\j} \times \unit{r}_3 \right ) - 5 \left(\ve{\j}\cdot \unit{r}_3 \right )^2 \left ( \ve{e}\times \unit{r}_3 \right ) + 35 \left ( \ve{e} \cdot \unit{r}_3 \right )^2 \left ( \ve{e} \times \unit{r}_3 \right ) \Biggl ]; \\
&\nonumber  \ve{f}_{\dot{\ve{e}},\mathrm{hex}}(\ve{e},\ve{\j},\unit{r}_3) \equiv \frac{15}{16} \Biggl [ 7 \left \{ 21 \left (\ve{e} \cdot \unit{r}_3 \right )^3 \left ( \ve{\j} \times \unit{r}_3 \right ) - 7 \left ( \ve{e} \cdot \unit{r}_3 \right )^2 \left ( \ve{\j} \cdot \unit{r}_3 \right ) \left ( \ve{e} \times \unit{r}_3 \right ) - 7 \left (\ve{e} \cdot \unit{r}_3 \right ) \left ( \ve{\j} \cdot \unit{r}_3 \right )^2 \left ( \ve{\j} \times \unit{r}_3 \right ) + \left ( \ve{\j} \cdot \unit{r}_3 \right )^3 \left ( \ve{e} \times \unit{r}_3 \right ) \right \}  \\
&\quad  + 7 \left ( \ve{e} \cdot \unit{r}_3 \right ) \left \{ \left ( \ve{\j} \times \unit{r}_3 \right ) - 10 \left ( \ve{e} \cdot \unit{r}_3 \right ) \left (\ve{\j} \times \ve{e} \right ) -10 e^2 \left (\ve{\j} \times \unit{r}_3 \right ) \right \} - \left (3-10e^2 \right ) \left ( \ve{\j} \cdot \unit{r}_3 \right )\left ( \ve{e} \times \unit{r}_3 \right ) + 10 \left (\ve{\j} \cdot \unit{r}_3\right )^2 \left (\ve{\j} \times \ve{e} \right ) - 2 \left (1-8 e^2 \right ) \left (\ve{\j} \times \ve{e} \right ) \Biggl ]; \\ 
& \ve{f}_{\dot{\ve{\j}},\mathrm{hex}}(\ve{e},\ve{\j},\unit{r}_3) \equiv \frac{15}{16} \Biggl [ 7 \left \{ 1 - 10e^2 + 21 \left ( \ve{e} \cdot \unit{r}_3 \right )^2 - 7 \left ( \ve{\j} \cdot \unit{r}_3\right )^2 \right \}\left (\ve{e} \cdot \unit{r}_3\right )  \left ( \ve{e} \times \unit{r}_3\right ) + \left \{ -3 + 10 e^2 - 49 \left (\ve{e}\cdot \unit{r}_3\right )^2 + 7 \left ( \ve{\j} \cdot \unit{r}_3\right )^2 \right \}  \left ( \ve{\j}\cdot \unit{r}_3 \right ) \left ( \ve{\j} \times \unit{r}_3 \right )\Biggl ]; \\
\nonumber & \ve{f}_{\dot{\ve{e}},\mathrm{hex,cross}}(\ve{e}_i,\ve{\j}_i,\ve{e}_{3-i},\ve{\j}_{3-i},\unit{r}_3) \equiv \frac{3}{16}  \Biggl [ -20 \left ( \ve{\j}_i \cdot \ve{e}_{3-i} \right ) \left (\ve{e}_i \times \ve{e}_{3-i} \right ) + 4 \left ( \ve{\j}_i \cdot \ve{\j}_{3-i} \right ) \left (\ve{e}_i \times  \ve{\j}_{3-i} \right ) + 10 \left (6e_{3-i}^2-1 \right ) \left (\ve{\j}_i \cdot \unit{r}_3 \right ) \left ( \ve{e}_i \times \unit{r}_3 \right ) \\
&\nonumber - 20 \left ( \ve{\j}_{3-i} \cdot \unit{r}_3 \right ) \left [ \left ( \ve{\j}_i \cdot \unit{r}_3 \right ) \left ( \ve{e}_i \times \ve{\j}_{3-i} \right ) + \left ( \ve{\j}_i \cdot \ve{\j}_{3-i} \right ) \left ( \ve{e}_i \times \unit{r}_3 \right ) \right ] + 100 \left ( \ve{e}_{3-i} \cdot \unit{r}_3 \right ) \left [ \left ( \ve{\j}_i \cdot \unit{r}_3 \right ) \left ( \ve{e}_i \times \ve{e}_{3-i} \right ) + \left ( \ve{\j}_i \cdot \ve{e}_{3-i} \right ) \left ( \ve{e}_i \times \unit{r}_3 \right ) \right ] \\
&\nonumber + 70 \left ( \ve{\j}_i \cdot \unit{r}_3 \right ) \left ( \ve{\j}_{3-i} \cdot \unit{r}_3 \right )^2 \left ( \ve{e}_i \times \unit{r}_3 \right ) -350 \left ( \ve{\j}_i \cdot \unit{r}_3 \right ) \left ( \ve{e}_{3-i} \cdot \unit{r}_3 \right )^2 \left ( \ve{e}_i \times \unit{r}_3 \right ) -12 \left ( \ve{\j}_i \times \ve{e}_i \right ) + 72 e_{3-i}^2 \left ( \ve{\j}_i \times \ve{e}_i \right ) + 100 \left ( \ve{e}_i \cdot \ve{e}_{3-i} \right ) \left ( \ve{\j}_i \times \ve{e}_{3-i} \right ) \\
&\nonumber -20 \left ( \ve{e}_i \cdot \ve{\j}_{3-i} \right ) \left ( \ve{\j}_i \times \ve{\j}_{3-i} \right ) + 50 \left ( \ve{e}_i \cdot \unit{r}_3 \right ) \left ( \ve{\j}_i \times \unit{r}_3 \right ) + 60 \left ( \ve{\j}_{3-i} \cdot \unit{r}_3 \right )^2 \left ( \ve{\j}_i \times \ve{e}_i \right ) - 300 e_{3-i}^2 \left (\ve{e}_i \cdot \unit{r}_3 \right ) \left ( \ve{\j}_i \times \unit{r}_3 \right ) - 300 \left ( \ve{e}_{3-i} \cdot \unit{r}_3 \right )^2 \left ( \ve{\j}_i \times \ve{e}_i \right ) \\
&\nonumber -500 \left ( \ve{e}_{3-i} \cdot \unit{r}_3 \right ) \left [ \left ( \ve{e}_i \cdot \unit{r}_3 \right ) \left ( \ve{\j}_i \times \ve{e}_{3-i} \right ) + \left ( \ve{e}_i \cdot \ve{e}_{3-i} \right ) \left (\ve{\j}_i \times \unit{r}_3 \right ) \right ] + 100 \left ( \ve{\j}_{3-i} \cdot \unit{r}_3 \right ) \left [ \left ( \ve{e}_i \cdot \unit{r}_3\right ) \left ( \ve{\j}_i \times \ve{\j}_{3-i} \right ) + \left ( \ve{e}_i \cdot \ve{\j}_{3-i} \right ) \left ( \ve{\j}_i \times \unit{r}_3 \right ) \right ] \\
&\nonumber + 1750 \left ( \ve{e}_i \cdot \unit{r}_3 \right ) \left ( \ve{e}_{3-i} \cdot \unit{r}_3 \right )^2 \left ( \ve{\j}_i \times \unit{r}_3 \right ) - 350 \left ( \ve{e}_i \cdot \unit{r}_3 \right ) \left ( \ve{\j}_{3-i} \cdot \unit{r}_3 \right )^2 \left ( \ve{\j}_i \times \unit{r}_3 \right ) \Biggl ]; \\
& \ve{f}_{\dot{\ve{j}},\mathrm{hex,cross}}(\ve{e}_i,\ve{\j}_i,\ve{e}_{3-i},\ve{\j}_{3-i},\unit{r}_3) \equiv \frac{3}{16} \Biggl [ 100 \left ( \ve{e}_i \cdot \ve{e}_{3-i} \right ) \left ( \ve{e}_i \times \ve{e}_{3-i} \right ) - 20 \left ( \ve{e}_i \cdot \ve{\j}_{3-i} \right ) \left ( \ve{e}_i \times \ve{\j}_{3-i} \right ) + 50 \left ( \ve{e}_i \cdot \unit{r}_3 \right ) \left ( \ve{e}_i \times \unit{r}_3 \right ) \\
&\nonumber - 300 e_{3-i}^2 \left ( \ve{e}_i \cdot \unit{r}_3 \right ) \left ( \ve{e}_i \times \unit{r}_3\right ) - 500 \left ( \ve{e}_i \cdot \unit{r}_3 \right ) \left [ \left ( \ve{e}_i \cdot \unit{r}_3 \right ) \left ( \ve{e}_i \times \ve{e}_{3-i} \right ) + \left ( \ve{e}_i \cdot \ve{e}_{3-i} \right ) \left ( \ve{e}_i \times \unit{r}_3 \right ) \right ] + 100 \left ( \ve{\j}_{3-i} \cdot \unit{r}_3 \right ) \left [ \left ( \ve{e}_i \cdot \unit{r}_3 \right ) \left (\ve{e}_i \times \ve{\j}_{3-i} \right ) \right. \\
&\nonumber \quad \left. + \left ( \ve{e}_i \cdot \ve{\j}_{3-i} \right ) \left ( \ve{e}_i \times \unit{r}_3 \right ) \right ] + 1750 \left ( \ve{e}_i \cdot \unit{r}_3 \right ) \left ( \ve{e}_{3-i} \cdot \unit{r}_3 \right )^2 \left ( \ve{e}_i \times \unit{r}_3 \right ) - 350 \left ( \ve{e}_i \cdot \unit{r}_3 \right ) \left ( \ve{\j}_{3-i} \cdot \unit{r}_3 \right )^2 \left ( \ve{e}_i \times \unit{r}_3 \right ) -20 \left ( \ve{\j}_i \cdot \ve{e}_{3-i} \right ) \left ( \ve{\j}_i \times \ve{e}_{3-i} \right ) \\
&\nonumber + 4 \left ( \ve{\j}_i \cdot \ve{\j}_{3-i} \right ) \left ( \ve{\j}_i \times \ve{\j}_{3-i} \right ) + 10 \left ( 6 e_{3-i}^2-1 \right ) \left ( \ve{\j}_i \cdot \unit{r}_3 \right ) \left ( \ve{\j}_i \times \unit{r}_3 \right ) - 20 \left ( \ve{\j}_{3-i} \cdot \unit{r}_3 \right ) \left [ \left ( \ve{\j}_i \cdot \unit{r}_3 \right ) \left ( \ve{\j}_i \times \ve{\j}_{3-i} \right ) + \left ( \ve{\j}_i \cdot \ve{\j}_{3-i} \right ) \left ( \ve{\j}_i \times \unit{r}_3 \right ) \right ] \\
&\nonumber + 100 \left ( \ve{e}_{3-i} \cdot \unit{r}_3\right ) \left [ \left ( \ve{\j}_i \cdot \unit{r}_3 \right ) \left ( \ve{\j}_i \times \ve{e}_{3-i} \right ) + \left ( \ve{\j}_i \cdot \ve{e}_{3-i} \right ) \left ( \ve{\j}_i \times \unit{r}_3 \right ) \right ] + 70 \left(\ve{\j}_i \cdot \unit{r}_3 \right ) \left ( \ve{\j}_{3-i} \cdot \unit{r}_3 \right)^2 \left ( \ve{\j}_i \times \unit{r}_3 \right ) - 350 \left (\ve{\j}_i \cdot \unit{r}_3 \right ) \left ( \ve{e}_{3-i} \cdot \unit{r}_3 \right )^2 \left ( \ve{\j}_i \times \unit{r}_3 \right ) \Biggl ]; \\
&\ve{f}_{\ddot{\ve{r}}_3} (m,m',a,\ve{e},\ve{\j},\ve{r}_3) =  \frac{m m'}{m+m'} \left ( \frac{a}{r_3} \right )^2 \frac{1}{4} \left [ -3 \left(1-6e^2\right ) \unit{r}_3 + 30 \left(\ve{e} \cdot \unit{r}_3 \right ) \ve{e} - 75 \left (\ve{e} \cdot \unit{r}_3\right)^2 \unit{r}_3 -6 \left(\ve{\j}\cdot \unit{r}_3 \right )\ve{\j} + 15 \left ( \ve{\j} \cdot \unit{r}_3\right)^2 \unit{r}_3 \right ].
\end{align}
\end{subequations}
Other (dimensionless) parameters appearing in \eq~(\ref{eq:eom}) are defined according to
\begin{subequations}
\begin{align}
\label{eq:epssa}
\epssa{i} &\equiv \left [ \frac{M_{3-i}^2}{M_i M} \left ( \frac{a_i}{\qper} \right )^3 \left (1+\eper \right)^{-3} \right ]^{1/2}; \\
\label{eq:epsoct}
\epsoct{i} &\equiv \frac{ \left  | m_{i,\mathrm{A}} - m_{i,\mathrm{B}} \right |}{M_i} \frac{a_i}{\qper} \frac{1}{1+\eper}; \\
\label{eq:epshex}
\epshex{i} &\equiv \frac{m_{i,\mathrm{A}}^2 - m_{i,\mathrm{A}} m_{i,\mathrm{B}} + m_{i,\mathrm{B}}^2}{M_i^2} \left ( \frac{a_i}{\qper} \right )^2 \frac{1}{(1+\eper)^2}; \\
\label{eq:epshexcross}
\epshexcross{i} &= \frac{m_{3-i,\mathrm{A}} m_{3-i,\mathrm{B}}}{M_{3-i}^2} \left ( \frac{a_{3-i}}{\qper} \right )^2 \frac{1}{(1+\eper)^2}.
\end{align}
\end{subequations}
Here, $m_{i,\mathrm{A}} = m_1$ and $m_{i,\mathrm{B}} = m_2$ if $i=1$, and $m_{i,\mathrm{A}} = m_3$ and $m_{i,\mathrm{B}} = m_4$ if $i=2$. For future convenience (\S~\ref{sect:an:an} below), we formulated the equations of motion for $\ve{e}_i$ and $\ve{\j}_i$  in terms of $\theta$, the true anomaly of the outer orbit, which is related to the physical time according to \eq~(\ref{eq:thetat}). Note that in the latter equation and in the inner-averaged approximation, $\eper$ and $a_3$ are allowed to vary and are determined by the equation for $\ve{r}_3$, \eq~(\ref{eq:eom:rdd}).

\subsection{Approximate analytic expressions for the eccentricity and angular-momentum changes}
\label{sect:an:an}
\subsubsection{Outer orbit}
\label{sect:an:an:out}
We first consider the backreaction of the outer orbit on the quadrupole moment of the inner two binaries. This effect is described by \Eq~(\ref{eq:eom:rdd}) to quadrupole expansion order (since the backreaction effect turns out to be small even at lowest order, we will not consider it at higher orders). We can get an approximate expression for the outer orbital changes ($\Delta a_3$, $\Delta e_3$, and $\Delta i_3$) by substituting the solution to \eq~(\ref{eq:eom:rdd}) in the absence of the perturbation terms $\propto \ve{f}_{\ddot{\ve{r}}_3}$ (i.e., the solution if $\ddot{\ve{r}}_3 = - GM/r_3^3 \ve{r}_3$, resulting in purely Keperian motion), into the perturbation terms and integrating the subsequent expressions over the outer orbit. Let the perturbation term to the Keplerian acceleration be denoted as 
\begin{align}
\ve{f}_3 \equiv  \frac{GM}{r_3^2} \ve{f}_{\ddot{\ve{r}}_3} (m_1,m_2,a_1,\ve{e}_1,\ve{\j}_1,\ve{r}_3) + \frac{GM}{r_3^2} \ve{f}_{\ddot{\ve{r}}_3} (m_3,m_4,a_2,\ve{e}_2,\ve{\j}_2,\ve{r}_3).
\end{align}
The changes to the outer semimajor axis, eccentricity vector, and (specific) angular-momentum vector ($\ve{h}_3 \equiv \ve{r}_3 \times \dot{\ve{r}}_3$) can then be found according to (e.g., \citealt{2006epbm.book.....E}, appendix C)
\begin{subequations}
\begin{align}
\frac{\Delta a_3}{a_3} &= \int_{-L}^{L} \mathrm{d} \theta \, \frac{-2a_3}{GM_1 M_2} \left (-  \dot{\ve{r}}_3 \cdot \ve{f}_3 \right ) \frac{\mathrm{d} t}{\mathrm{d} \theta} = 0; \\
\Delta \ve{e}_3 &= \int_{-L}^L \mathrm{d} \theta\,  \frac{1}{GM} \left [2 \ve{r}_3 \left (\dot{\ve{r}}_3 \cdot \ve{f}_3 \right ) - \ve{f}_3 \left (\ve{r}_3 \cdot \dot{\ve{r}}_3 \right ) - \dot{\ve{r}}_3 \left ( \ve{r}_3 \cdot \ve{f}_3 \right ) \right ]  \frac{\mathrm{d} t}{\mathrm{d} \theta} = \ve{f}_{\Delta \ve{e}_3} (m_1,m_2,a_1,\ve{e}_1,\ve{\j}_1,\eper) + \ve{f}_{\Delta \ve{e}_3} (m_3,m_4,a_2,\ve{e}_2,\ve{\j}_2,\eper); \\
\Delta \ve{h}_3/h_3 &= \int_{-L}^L \mathrm{d} \theta \, \frac{1}{h_3} \left ( \ve{r}_3 \times \ve{f}_3 \right ) \frac{\mathrm{d} t}{\mathrm{d} \theta} = \ve{f}_{\Delta \ve{h}_3} (m_1,m_2,a_1,\ve{e}_1,\ve{\j}_1,\eper) + \ve{f}_{\Delta \ve{h}_3} (m_3,m_4,a_2,\ve{e}_2,\ve{\j}_2,\eper).
\end{align}
\end{subequations}
Here, we defined the additional expressions
\begin{subequations}
\begin{align}
\nonumber &\ve{f}_{\Delta \ve{e}_3} (m,m',a,\ve{e},\ve{\j},\eper) \equiv \frac{m m'}{(m+m')^2} \left ( \frac{a}{\qper} \right )^2 \Biggl [ \frac{(\eper-1)^{5/2} \sqrt{\eper+1}}{\eper^3 }(5 e_x e_y-\j_x \j_y)\, \unit{x} + \frac{3}{32 (\eper+1)^2} \Biggl \{ \frac{8 \sqrt{\eper^2-1}}{3 \eper^3} \left(4 \eper^4 \left( 1 + 4 e_x^2 - e_y^2 - 6 e_z^2 \right. \right.  \\
\nonumber &\quad \left. \left.  - 2 \j_x^2 - \j_y^2\right)+\eper^2 \left( 2 - 17 e_x^2 + 23 e_y^2 - 12 e_z^2 + \j_x^2 - 7 \j_y^2\right)+2 \left(5 e_x^2-5 e_y^2-\j_x^2+\j_y^2\right)\right) + 8 \eper L \left(2 + 3 e_x^2 + 3 e_y^2 - 12 e_z^2 - 3 \j_x^2 - 3 \j_y^2\right)\Biggl \} \, \unit{y} \\
&\quad -\frac{ \sqrt{\eper^2-1} \left(2 \eper^2+1\right)+3 \eper^2 L }{2 \eper (\eper+1)^2} (5 e_y e_z-\j_y \j_z) \, \unit{z} \Biggl ]; \\
\nonumber &\ve{f}_{\Delta \ve{h}_3} (m,m',a,\ve{e},\ve{\j},\eper) \equiv\frac{m m'}{(m+m')^2} \left ( \frac{a}{\qper} \right )^2  \Biggl  [\frac{\sqrt{1-\frac{1}{\eper^2}} \left(2 \eper^2+1\right)+3 \eper L}{2 \eper (\eper+1)^2} (5 e_y e_z-\j_y \j_z) \,\unit{x} - \frac{ \sqrt{1-\frac{1}{\eper^2}} \left(4 \eper^2-1\right)+3 \eper L }{2 \eper (\eper+1)^2} (5 e_x e_z-\j_x \j_z) \,\unit{y} \\
&\quad + \frac{ (\eper-1)^{3/2}}{\eper^2 \sqrt{\eper+1}} (5 e_x e_y-\j_x \j_y) \, \unit{z} \Biggl ].
\end{align}
\end{subequations}

Since $\unit{e}_3 = \unit{x}$ initially, for small perturbations, the scalar eccentricity change is given by
\begin{align}
\Delta e_3 \simeq \unit{e}_3 \cdot \Delta \ve{e}_3 =  \frac{(\eper-1)^{5/2} \sqrt{\eper+1}}{\eper^3 } \left [ \frac{m_1 m_2}{M_1^2} \left ( \frac{a_1}{\qper} \right )^2 \left (5e_{1,x} e_{1,y} - \j_{1,x} \j_{1,y} \right )  + \frac{m_3 m_4}{M_2^2} \left ( \frac{a_2}{\qper} \right )^2 \left (5e_{2,x} e_{2,y} - \j_{2,x} \j_{2,y} \right )  \right ].
\end{align}
Note that $\Delta e_3 = 0$ if $\eper=1$ (parabolic orbits), and becomes independent of $\eper$ as $\eper \gg 1$. 

The inclination change, $\Delta i_3$, is obtained from the new $\ve{h}_3' = h_3 \unit{z} + \Delta \ve{h}_3$ and noting that the inclination is measured with respect to the $z$-axis, giving
\begin{align}
&\nonumber \cos \Delta i_3 = \frac{1 + \frac{\ve{h}_3 \cdot \unit{z}}{h_3}}{\left |\left| \unit{z} + \frac{\Delta \ve{h}_3}{h_3} \right| \right|} = 2 \eper \Biggl [ a_1^2 (\eper-1)^{3/2} m_1 m_2 M_2^2 (5 e_{1,x} e_{1,y}-\j_{1,x} \j_{1,y})+a_2^2 (\eper-1)^{3/2} m_3 m_4 M_1^2 (5 e_{2,x} e_{2,y}-\j_{2,x} \j_{2,y})+\eper^2 \sqrt{\eper+1} M_1^2 M_2^2 Q^2 \Biggl ] \\
\nonumber &\quad \times \left(1-\eper^2\right)^{-3/2} \Biggl [4 Q^3 \left(1-\eper^2 \right)^3 \left(\sqrt{\frac{(\eper-1)^3 \eper^2}{Q^3}} \left(a_1^2 m_1 m_2 M_2^2 (5 e_{1,x} e_{1,y}-\j_{1,x} \j_{1,y})+a_2^2 m_3 m_4 M_1^2 (5 e_{2,x} e_{2,y}-\j_{2,x} \j_{2,y})\right) \right. \\
\nonumber &\quad \left. +\eper^3 M_1^2 M_2^2 \sqrt{(\eper+1) Q}\right)^2+ (1-\eper)^3 \eper^4 \left(\sqrt{1-\frac{1}{\eper^2}} \left(4 \eper^2-1\right)+3 \eper L\right)^2 \left(a_1^2 m_1 m_2 M_2^2 (5 e_{1,x} e_{1,z}-\j_{1,x} \j_{1,z}) \right. \\
\nonumber &\quad \left. +a_2^2 m_3 m_4 M_1^2 (5 e_{2,x} e_{2,z}-\j_{2,x} \j_{2,z})\right)^2+(1-\eper)^3 \eper^4 \left(\sqrt{1-\frac{1}{\eper^2}} \left(2 \eper^2+1\right)+3 \eper L)\right)^2 \left(a_1^2 m_1 m_2 M_2^2 (5 e_{1,y} e_{1,z}-\j_{1,y} \j_{1,z}) \right. \\
&\quad \left. +a_2^2 m_3 m_4 M_1^2 (5 e_{2,y} e_{2,z}-\j_{2,y} \j_{2,z})\right)^2 \Biggl ]^{-1/2}.
\end{align}
Here, we used that the initial $h_3 = \sqrt{GM\qper(1+\eper)}$. 

From these equations for $\Delta e_3$ and $\Delta i_3$, it is clear that the backreaction effects scale with $(a_i/\qper)^2$ and so are typically small. This is also borne out by numerical simulations below (\S~\ref{sect:num}).

\subsubsection{Inner orbits}
\label{sect:an:an:in}
We can obtain approximate expressions for the scalar eccentricity change of orbit $i$ ($i\in\{1,2\}$) by integrating the equations of motion, \eq~(\ref{eq:eom}), over $\theta$ assuming that all orbits (including the outer orbit) are static (i.e., constant $\ve{e}_i$ and $\ve{\j}_i$). The result is
\begin{align}
\label{eq:Delta_e_i_FO}
\Delta e_i = \Delta e_{i,\mathrm{quad}} + \Delta e_{i,\mathrm{oct}} + \Delta e_{i,\mathrm{hex}} + \Delta e_{i,\mathrm{hex,cross}},
\end{align}
where
\begin{subequations}
\begin{align}
&\Delta e_{i,\mathrm{quad}} = \epssa{i} \frac{5}{2e_i \eper} \left(\sqrt{1-\frac{1}{\eper^2}} \left(2 e_{i,x} e_{i,y} \left(\eper^2-1\right) \j_{i,z}+e_{i,x} e_{i,z} \left(1-4 \eper^2\right) \j_{i,y}+e_{i,y} e_{i,z} \left(2 \eper^2+1\right) \j_{i,x}\right)+3 e_{i,z} \eper L (e_{i,y} \j_{i,x}-e_{i,x} \j_{i,y})\right); \\
\nonumber &\Delta e_{i,\mathrm{oct}} = \epssa{i} \epsoct{i} \frac{5}{32e_i \eper^2} \Biggl [ 3 \eper^3 L \left(e_{i,x}^2 (3 e_{i,y} \j_{i,z}-73 e_{i,z} \j_{i,y})+10 e_{i,x} \j_{i,x} (7 e_{i,y} e_{i,z}+\j_{i,y} \j_{i,z}) \right.  \\
\nonumber &\quad  \left. +e_{i,z} \j_{i,y} \left(-3 e_{i,y}^2+5 \j_{i,x}^2+5 \j_{i,y}^2-4\right)+e_{i,y} \j_{i,z} \left(3 e_{i,y}^2-15 \j_{i,x}^2-5 \j_{i,y}^2+4\right)-32 e_{i,y} e_{i,z}^2 \j_{i,z}+32 e_{i,z}^3 \j_{i,y}\right)   \\
\nonumber&\quad  +\sqrt{1-\frac{1}{\eper^2}} \left(-e_{i,x}^2 \left(e_{i,z} \left(160 \eper^4+45 \eper^2+14\right) \j_{i,y}-3 e_{i,y} \left(16 \eper^4-27 \eper^2+14\right) \j_{i,z}\right)  \right. \\
\nonumber&\quad \left. +2 e_{i,x} \left(8 \eper^4+9 \eper^2-2\right) \j_{i,x} (7 e_{i,y} e_{i,z}+\j_{i,y} \j_{i,z})+e_{i,y}^3 \left(-8 \eper^4+31 \eper^2-14\right) \j_{i,z}+e_{i,y}^2 e_{i,z} \left(8 \eper^4-31 \eper^2+14\right) \j_{i,y} \right. \\
\nonumber&\quad \left. -e_{i,y} \j_{i,z} \left(8 \eper^4 \left(8 e_{i,z}^2+4 \j_{i,x}^2+\j_{i,y}^2-1\right)+\eper^2 \left(32 e_{i,z}^2+11 \j_{i,x}^2+9 \j_{i,y}^2-4\right)+2 \left(\j_{i,x}^2-\j_{i,y}^2\right)\right) \right. \\
&\quad \left. +e_{i,z} \j_{i,y} \left(8 \eper^4 \left(8 e_{i,z}^2+2 \j_{i,x}^2+\j_{i,y}^2-1\right)+\eper^2 \left(32 e_{i,z}^2-7 \j_{i,x}^2+9 \j_{i,y}^2-4\right)+6 \j_{i,x}^2-2 \j_{i,y}^2\right)\right) \Biggl ]; \\
&\nonumber \Delta e_{i,\mathrm{hex}} = \epssa{i} \epshex{i} \frac{7}{128 e_i \eper^3 } \Biggl [ 15 \eper^3 L \left(e_{i,y} \left(e_{i,z} \j_{i,x} \left(e_{i,x}^2 \left(129 \eper^2+46\right)+\eper^2 \left(-21 \j_{i,x}^2+21 \j_{i,y}^2+6\right)-2 \left(7 \j_{i,x}^2+7 \j_{i,y}^2-4\right)\right) \right. \right. \\
\nonumber &\quad \left. \left. +2 e_{i,x} \j_{i,z} \left(3 \eper^2 \left(e_{i,x}^2-14 \j_{i,x}^2+2\right)+14 \left(\j_{i,y}^2-\j_{i,x}^2\right)\right)-120 e_{i,x} e_{i,z}^2 \eper^2 \j_{i,z}-20 e_{i,z}^3 \left(3 \eper^2+4\right) \j_{i,x}\right) \right. \\
\nonumber &\quad \left. +e_{i,x} \j_{i,y} \left(e_{i,z} \left(e_{i,x}^2 \left(-\left(135 \eper^2+46\right)\right)+3 \eper^2 \left(21 \j_{i,x}^2+7 \j_{i,y}^2-6\right)+2 \left(7 \j_{i,x}^2+7 \j_{i,y}^2-4\right)\right)+14 e_{i,x} \left(3 \eper^2+2\right) \j_{i,x} \j_{i,z} \right. \right. \\
\nonumber &\quad \left. \left. +20 e_{i,z}^3 \left(9 \eper^2+4\right)\right)+e_{i,y}^3 \left(6 e_{i,x} \eper^2 \j_{i,z}+e_{i,z} \left(3 \eper^2+46\right) \j_{i,x}\right)-e_{i,y}^2 \j_{i,y} \left(e_{i,x} e_{i,z} \left(9 \eper^2+46\right)+14 \left(3 \eper^2+2\right) \j_{i,x} \j_{i,z}\right)\right) \\
\nonumber &\quad +\sqrt{1-\frac{1}{\eper^2}} \left(e_{i,x}^3 \left(6 e_{i,y} \left(32 \eper^6-63 \eper^4+70 \eper^2-24\right) \j_{i,z}+e_{i,z} \left(-1024 \eper^6-1751 \eper^4+24 \eper^2+36\right) \j_{i,y}\right) \right. \\
\nonumber &\quad \left. +e_{i,x}^2 \j_{i,x} \left(e_{i,y} e_{i,z} \left(832 \eper^6+2129 \eper^4-444 \eper^2+108\right)+2 \left(128 \eper^6+421 \eper^4-36 \eper^2+12\right) \j_{i,y} \j_{i,z}\right) \right. \\
\nonumber &\quad \left. +e_{i,x} \left(-6 e_{i,y}^3 \left(16 \eper^6-81 \eper^4+74 \eper^2-24\right) \j_{i,z}+e_{i,y}^2 e_{i,z} \left(128 \eper^6-1049 \eper^4+204 \eper^2-108\right) \j_{i,y} \right. \right. \\
\nonumber &\quad \left. \left. -4 e_{i,y} \j_{i,z} \left(8 \eper^6 \left(30 e_{i,z}^2+20 \j_{i,x}^2+\j_{i,y}^2-3\right)+\eper^4 \left(270 e_{i,z}^2+269 \j_{i,x}^2-80 \j_{i,y}^2-27\right)-3 \eper^2 \left(20 e_{i,z}^2+\j_{i,x}^2+13 \j_{i,y}^2-2\right)-6 \j_{i,x}^2+6 \j_{i,y}^2\right) \right. \right. \\
\nonumber &\quad \left. \left. +e_{i,z} \j_{i,y} \left(128 \eper^6 \left(10 e_{i,z}^2+4 \j_{i,x}^2+\j_{i,y}^2-1\right)+\eper^4 \left(2740 e_{i,z}^2+655 \j_{i,x}^2+421 \j_{i,y}^2-274\right)-12 \eper^2 \left(10 e_{i,z}^2-2 \j_{i,x}^2+3 \j_{i,y}^2-1\right) \right. \right. \right. \\
\nonumber &\quad \left. \left. \left. +12 \left(\j_{i,y}^2-3 \j_{i,x}^2\right)\right)\right)-e_{i,y} \j_{i,x} \left(e_{i,y}^2 e_{i,z} \left(32 \eper^6-563 \eper^4-240 \eper^2+36\right)+2 e_{i,y} \left(128 \eper^6+421 \eper^4-36 \eper^2+12\right) \j_{i,y} \j_{i,z} \right. \right. \\
\nonumber &\quad \left. \left. +e_{i,z} \left(32 \eper^6 \left(10 e_{i,z}^2+4 \j_{i,x}^2-5 \j_{i,y}^2-1\right)+\eper^4 \left(1660 e_{i,z}^2+421 \j_{i,x}^2-101 \j_{i,y}^2-166\right)+12 \eper^2 \left(10 e_{i,z}^2-3 \j_{i,x}^2+16 \j_{i,y}^2-1\right) \right. \right. \right. \\
&\quad \left. \left. \left. +12 \left(\j_{i,x}^2-3 \j_{i,y}^2\right)\right)\right)\right) \Biggl ]; \\
\nonumber &\Delta e_{i,\mathrm{hex,cross}} = \epssa{i} \epshexcross{i} \frac{5}{64 e_i \eper^3} \Biggl [ 3 \left(-6 \left(5 e_{3-i,x} \left(2 e_{3-i,z} \left(11 \eper^2+4\right) \j_{i,y}+e_{3-i,y} \left(3 \eper^2+2\right) \j_{i,z}\right)-\j_{3-i,x} \left(\left(3 \eper^2+2\right) \j_{i,z} \j_{3-i,y} \right. \right. \right. \\
\nonumber &\quad \left. \left. \left. +2 \left(11 \eper^2+4\right) \j_{i,y} \j_{3-i,z}\right)\right) e_{i,x}^2+e_{i,z} \left(-5 \left(105 \eper^2+34\right) \j_{i,y} e_{3-i,x}^2+30 e_{3-i,y} \left(3 \eper^2+2\right) \j_{i,x} e_{3-i,x}+60 e_{3-i,y} e_{3-i,z} \left(\eper^2+4\right) \j_{i,z} \right. \right. \\
\nonumber &\quad \left. \left. -6 \j_{3-i,y} \left(\left(3 \eper^2+2\right) \j_{i,x} \j_{3-i,x}+2 \left(\eper^2+4\right) \j_{i,z} \j_{3-i,z}\right)+\j_{i,y} \left(5 \left(3 \eper^2-22\right) e_{3-i,y}^2-90 \eper^2+213 \eper^2 \j_{3-i,x}^2+82 \j_{3-i,x}^2+105 \eper^2 \j_{3-i,y}^2 \right. \right. \right. \\
\nonumber &\quad \left. \left. \left. +70 \j_{3-i,y}^2-48 \eper^2 \j_{3-i,z}^2-32 \j_{3-i,z}^2+20 e_{3-i,z}^2 \left(39 \eper^2+20\right)-40\right)\right) e_{i,x}+12 e_{i,z}^2 \left(-5 e_{3-i,y} e_{3-i,z} \left(\eper^2+4\right) \j_{i,x}+\left(\eper^2+4\right) \j_{3-i,y} \j_{3-i,z} \j_{i,x} \right. \right. \\
\nonumber &\quad \left. \left. +5 e_{3-i,x} e_{3-i,z} \left(11 \eper^2+4\right) \j_{i,y}-\left(11 \eper^2+4\right) \j_{i,y} \j_{3-i,x} \j_{3-i,z}\right)+6 e_{i,y}^2 \left(5 e_{3-i,y} \left(2 e_{3-i,z} \left(\eper^2+4\right) \j_{i,x}+e_{3-i,x} \left(3 \eper^2+2\right) \j_{i,z}\right) \right. \right. \\
\nonumber &\quad \left. \left. - \j_{3-i,y} \left(\left(3 \eper^2+2\right) \j_{i,z} \j_{3-i,x}+2 \left(\eper^2+4\right) \j_{i,x} \j_{3-i,z}\right)\right)+e_{i,y} \left(12 e_{i,x} \left(5 \left(3 \j_{i,z} \eper^2+\j_{i,z}\right) e_{3-i,x}^2+5 e_{3-i,z} \left(11 \eper^2+4\right) \j_{i,x} e_{3-i,x} \right. \right. \right. \\
\nonumber &\quad \left. \left. \left. - 9 \eper^2 \j_{i,z} \j_{3-i,x}^2-\j_{i,z} \j_{3-i,x}^2-6 \eper^2 \j_{i,z} \j_{3-i,y}^2+\j_{i,z} \j_{3-i,y}^2-5 e_{3-i,y} e_{3-i,z} \left(\eper^2+4\right) \j_{i,y}-5 e_{3-i,y}^2 \j_{i,z}-30 e_{3-i,z}^2 \eper^2 \j_{i,z}+5 \eper^2 \j_{i,z} \right. \right. \right. \\
\nonumber &\quad \left. \left. \left. - 11 \eper^2 \j_{i,x} \j_{3-i,x} \j_{3-i,z}-4 \j_{i,x} \j_{3-i,x} \j_{3-i,z}+\eper^2 \j_{i,y} \j_{3-i,y} \j_{3-i,z}+4 \j_{i,y} \j_{3-i,y} \j_{3-i,z}\right)+e_{i,z} \left(-30 e_{3-i,x} \left(e_{3-i,y} \left(3 \eper^2+2\right) \j_{i,y} \right. \right. \right. \right. \\
\nonumber &\quad \left. \left. \left. \left. +2 e_{3-i,z} \left(11 \eper^2+4\right) \j_{i,z}\right)+6 \j_{3-i,x} \left(\left(3 \eper^2+2\right) \j_{i,y} \j_{3-i,y}+2 \left(11 \eper^2+4\right) \j_{i,z} \j_{3-i,z}\right)+\j_{i,x} \left(5 \left(69 \eper^2+22\right) e_{3-i,x}^2-400 e_{3-i,z}^2 \right. \right. \right. \right. \\
\nonumber &\quad \left. \left. \left. \left. -420 e_{3-i,z}^2 \eper^2+30 \eper^2-105 \eper^2 \j_{3-i,x}^2-70 \j_{3-i,x}^2-33 \eper^2 \j_{3-i,y}^2-82 \j_{3-i,y}^2+48 \eper^2 \j_{3-i,z}^2+32 \j_{3-i,z}^2-5 e_{3-i,y}^2 \left(3 \eper^2-34\right)+40\right)\right)\right)\right)L \eper^3 \\
\nonumber &\quad +\sqrt{1-\frac{1}{\eper^2}} \Biggl \{ -6 \left(5 e_{3-i,x} \left(2 e_{3-i,z} \left(16 \eper^4+31 \eper^2-2\right) \j_{i,y} \eper^2+e_{3-i,y} \left(23 \eper^4-12 \eper^2+4\right) \j_{i,z}\right)-\j_{3-i,x} \left(2 \left(16 \eper^4+31 \eper^2-2\right) \j_{i,y} \j_{3-i,z} \eper^2 \right. \right.  \\
\nonumber &\quad  \left. \left. +\left(23 \eper^4-12 \eper^2+4\right) \j_{i,z} \j_{3-i,y}\right)\right) e_{i,x}^2+e_{i,z} \left(320 \j_{i,y} \j_{3-i,x}^2 \eper^6+128 \j_{i,y} \j_{3-i,y}^2 \eper^6-64 \j_{i,y} \j_{3-i,z}^2 \eper^6+1088 e_{3-i,z}^2 \j_{i,y} \eper^6-128 \j_{i,y} \eper^6 \right.  \\
\nonumber &\quad  \left. +577 \j_{i,y} \j_{3-i,x}^2 \eper^4+421 \j_{i,y} \j_{3-i,y}^2 \eper^4-176 \j_{i,y} \j_{3-i,z}^2 \eper^4+2524 e_{3-i,z}^2 \j_{i,y} \eper^4-274 \j_{i,y} \eper^4-138 \j_{i,x} \j_{3-i,x} \j_{3-i,y} \eper^4-156 \j_{i,z} \j_{3-i,y} \j_{3-i,z} \eper^4 \right.  \\
\nonumber &\quad  \left. -36 \j_{i,y} \j_{3-i,y}^2 \eper^2-72 e_{3-i,z}^2 \j_{i,y} \eper^2+12 \j_{i,y} \eper^2+60 e_{3-i,y} e_{3-i,z} \left(13 \eper^2+2\right) \j_{i,z} \eper^2+72 \j_{i,x} \j_{3-i,x} \j_{3-i,y} \eper^2-24 \j_{i,z} \j_{3-i,y} \j_{3-i,z} \eper^2 \right.  \\
\nonumber &\quad  \left. -12 \j_{i,y} \j_{3-i,x}^2+12 \j_{i,y} \j_{3-i,y}^2+30 e_{3-i,x} e_{3-i,y} \left(23 \eper^4-12 \eper^2+4\right) \j_{i,x}+e_{3-i,y}^2 \left(128 \eper^6-461 \eper^4+108 \eper^2-60\right) \j_{i,y} \right.  \\
\nonumber &\quad  \left. -e_{3-i,x}^2 \left(832 \eper^6+1241 \eper^4+72 \eper^2-60\right) \j_{i,y}-24 \j_{i,x} \j_{3-i,x} \j_{3-i,y}\right) e_{i,x}+12 e_{i,z}^2 \eper^2 \left(-5 e_{3-i,y} e_{3-i,z} \left(13 \eper^2+2\right) \j_{i,x}\right.  \\
\nonumber &\quad  \left. +\left(13 \eper^2+2\right) \j_{3-i,y} \j_{3-i,z} \j_{i,x}+5 e_{3-i,x} e_{3-i,z} \left(16 \eper^4+31 \eper^2-2\right) \j_{i,y}+\left(-16 \eper^4-31 \eper^2+2\right) \j_{i,y} \j_{3-i,x} \j_{3-i,z}\right)  \\
\nonumber &\quad  +6 e_{i,y}^2 \left(5 e_{3-i,y} \left(2 e_{3-i,z} \left(13 \eper^2+2\right) \j_{i,x} \eper^2+e_{3-i,x} \left(23 \eper^4-12 \eper^2+4\right) \j_{i,z}\right)-\j_{3-i,y} \left(2 \left(13 \eper^2+2\right) \j_{i,x} \j_{3-i,z} \eper^2 \right. \right.  \\
\nonumber &\quad  \left. \left. +\left(23 \eper^4-12 \eper^2+4\right) \j_{i,z} \j_{3-i,x}\right)\right)+e_{i,y} \left(12 e_{i,x} \left(-16 \j_{i,z} \j_{3-i,x}^2 \eper^6-8 \j_{i,z} \j_{3-i,y}^2 \eper^6-48 e_{3-i,z}^2 \j_{i,z} \eper^6+8 \j_{i,z} \eper^6-16 \j_{i,x} \j_{3-i,x} \j_{3-i,z} \eper^6 \right. \right.  \\
\nonumber &\quad  \left. \left. -13 \j_{i,z} \j_{3-i,x}^2 \eper^4-14 \j_{i,z} \j_{3-i,y}^2 \eper^4-54 e_{3-i,z}^2 \j_{i,z} \eper^4+9 \j_{i,z} \eper^4-31 \j_{i,x} \j_{3-i,x} \j_{3-i,z} \eper^4+13 \j_{i,y} \j_{3-i,y} \j_{3-i,z} \eper^4-3 \j_{i,z} \j_{3-i,x}^2 \eper^2 \right. \right.  \\
\nonumber &\quad  \left. \left. +9 \j_{i,z} \j_{3-i,y}^2 \eper^2+5 e_{3-i,x} e_{3-i,z} \left(16 \eper^4+31 \eper^2-2\right) \j_{i,x} \eper^2-5 e_{3-i,y} e_{3-i,z} \left(13 \eper^2+2\right) \j_{i,y} \eper^2+12 e_{3-i,z}^2 \j_{i,z} \eper^2-2 \j_{i,z} \eper^2 \right. \right.  \\
\nonumber &\quad  \left. \left. +2 \j_{i,x} \j_{3-i,x} \j_{3-i,z} \eper^2+2 \j_{i,y} \j_{3-i,y} \j_{3-i,z} \eper^2+2 \j_{i,z} \j_{3-i,x}^2-2 \j_{i,z} \j_{3-i,y}^2+e_{3-i,y}^2 \left(-8 \eper^6+16 \eper^4-33 \eper^2+10\right) \j_{i,z} \right. \right.  \\
\nonumber &\quad  \left. \left. +e_{3-i,x}^2 \left(32 \eper^6+11 \eper^4+27 \eper^2-10\right) \j_{i,z}\right)+e_{i,z} \left(-128 \j_{i,x} \j_{3-i,x}^2 \eper^6-32 \j_{i,x} \j_{3-i,y}^2 \eper^6+64 \j_{i,x} \j_{3-i,z}^2 \eper^6-512 e_{3-i,z}^2 \j_{i,x} \eper^6+32 \j_{i,x} \eper^6 \right. \right.  \\
\nonumber &\quad  \left. \left. +192 \j_{i,z} \j_{3-i,x} \j_{3-i,z} \eper^6-421 \j_{i,x} \j_{3-i,x}^2 \eper^4-253 \j_{i,x} \j_{3-i,y}^2 \eper^4+176 \j_{i,x} \j_{3-i,z}^2 \eper^4-1876 e_{3-i,z}^2 \j_{i,x} \eper^4+166 \j_{i,x} \eper^4 \right. \right.  \\
\nonumber &\quad \left. \left. +138 \j_{i,y} \j_{3-i,x} \j_{3-i,y} \eper^4+372 \j_{i,z} \j_{3-i,x} \j_{3-i,z} \eper^4+36 \j_{i,x} \j_{3-i,x}^2 \eper^2-72 \j_{i,x} \j_{3-i,y}^2 \eper^2-72 e_{3-i,z}^2 \j_{i,x} \eper^2+12 \j_{i,x} \eper^2-72 \j_{i,y} \j_{3-i,x} \j_{3-i,y} \eper^2 \right. \right.  \\
\nonumber &\quad \left. \left. -24 \j_{i,z} \j_{3-i,x} \j_{3-i,z} \eper^2-12 \j_{i,x} \j_{3-i,x}^2+12 \j_{i,x} \j_{3-i,y}^2+e_{3-i,y}^2 \left(-32 \eper^6+269 \eper^4+288 \eper^2-60\right) \j_{i,x} \right. \right.  \\
\nonumber &\quad \left. \left. +e_{3-i,x}^2 \left(448 \eper^6+1109 \eper^4-252 \eper^2+60\right) \j_{i,x}-30 e_{3-i,x} \left(2 e_{3-i,z} \left(16 \eper^4+31 \eper^2-2\right) \j_{i,z} \eper^2+e_{3-i,y} \left(23 \eper^4-12 \eper^2+4\right) \j_{i,y}\right) \right. \right.  \\
&\quad  \left. \left. +24 \j_{i,y} \j_{3-i,x} \j_{3-i,y}\right)\right)\Biggl \}\Biggl ].
\end{align}
\end{subequations}

Here, for simplicity, we neglected corrections due to changes of the orbits during the encounter, i.e., we restricted to terms of order $\epssa{i}$ and neglected terms of order $\epssa{i}^2$ and higher (see \citealt{2019MNRAS.487.5630H}). However, when comparing to numerical integrations in \S~\ref{sect:num:bin}, we do include the quadrupole-order term $\propto \epssa{i}^2$, where the corresponding expression was derived in \citet{2019MNRAS.487.5630H}.

\section{Numerical integrations}
\label{sect:num}
In this section, we carry out several numerical integrations to illustrate orbital changes in the two binaries for various parameters, and compare to the analytic expressions of \S~\ref{sect:an:an}. In \S~\ref{sect:num:out}, we focus on the backreaction of the outer orbit; in \S~\ref{sect:num:bin}, we consider series of integrations with varying properties of binary 2. An overview of the initial conditions adopted in these sections is given in Table~\ref{table:values}. We choose to restrict to systems with equal masses in binary 1, which is motivated by the fact that this eliminates the octuple-order terms (see \eq~\ref{eq:epsoct}), which would otherwise dominate the hexadecupole-order terms and thus decrease the importance of the hexadecupole-order cross term even further. 

Our numerical integrations are based on four-body calculations, as well as calculations based on the equations of motion averaged over the inner orbits (see \S~\ref{sect:an:eom}). The four-body integrations were carried out using the \textsc{IAS15} integrator within the \textsc{Rebound} package \citep{2012A&A...537A.128R,2015MNRAS.446.1424R}. We integrated the inner-averaged equations of motion using \textsc{odeint} from the \textsc{Python} \textsc{Scipy} library, with the relative and absolute tolerances set to $10^{-13}$. In both cases of the four-body and inner-averaged integrations, the integration time was set to $t_\mathrm{end}$ with periapsis passage (ignoring backreaction) occurring at $t_\mathrm{end}/2$, where
\begin{align}
\label{eq:tend}
t_\mathrm{end} = \frac{1}{n_3} \left [ -4 \, \mathrm{arctanh} \left ( \frac{ \left (\eper-1 \right ) \tan (\beta/2)}{\sqrt{\eper^2-1}} \right ) + \frac{2 \eper \sqrt{\eper^2-1} \sin\beta}{1 + \eper \cos \beta} \right ].
\end{align}
Here, $\beta = f_\theta \arccos(-1/\eper)$ indicates the fraction of the outer orbit true anomaly $\theta$ in the integrations compared to integrating from $t\rightarrow -\infty$ to $t\rightarrow \infty$. Specifically, $f_\theta$ corresponds to integrating over true anomaly $\theta$ from $-f_\theta \arccos(-1/\eper)$ to $f_\theta \arccos(-1/\eper)$, with $f_\theta=1$ corresponding to integrating from $t\rightarrow -\infty$ to $t\rightarrow \infty$. We have checked our results for convergence with respect to $f_\theta$.

Several \textsc{Python} scripts to carry out the four-body and inner-averaged integrations and to compute the analytic expressions are freely available\footnote{\href{https://github.com/hamers/flybys_bin}{https://github.com/hamers/flybys\_bin}}.

\begin{table}
\begin{tabular}{lccccccccccccccccccc}
\toprule
& $m_1$ & $m_2$ & $m_3$ & $m_4$ & $a_1$ & $a_2$ & $Q$ & $e_1$ & $e_2$ & $e_3$ & $i_1$ & $i_2$ & $\omega_1$ & $\omega_2$ & $\Omega_1$ & $\Omega_2$ & $\theta_1$ & $\theta_2$ & $f_\theta$ \\
\midrule
\F~\ref{fig:backreaction} & 10 & 10 & 5 & 5 & 1.0 & 1.5 & 20 & 0.9 & 0.5 & 1.5 & 90 & 0.01 & 45 & 0.01 & 0.01 & 0.01 & 0.01 & 0.01 & 0.98 \\
\F~\ref{fig:series_a2_low_e} & 10 & 10 & 5 & 5 & 1.0 & 0.5-3 & 20 & 0.1 & 0.4 & 1.5 & 90 & 0.01 & 45 & 0.01 & 0.01 & 0.01 & 0.01 & 0.01 & 0.98 \\
\F~\ref{fig:series_a2_low_e_compare_back} & 10 & 10 & 5 & 5 & 1.0 & 0.5-3 & 20 & 0.1 & 0.4 & 1.5 & 90 & 0.01 & 45 & 0.01 & 0.01 & 0.01 & 0.01 & 0.01 & 0.98 \\
\F~\ref{fig:series_a2_high_e} & 10 & 10 & 5 & 5 & 1.0 & 0.5-3 & 20 & 0.9 & 0.4 & 1.5 & 90 & 0.01 & 45 & 0.01 & 0.01 & 0.01 & 0.01 & 0.01 & 0.98 \\
\F~\ref{fig:series_a2_angle} & 10 & 10 & 5 & 5 & 1.0 & 0.5-3 & 20 & 0.9 & 0.4 & 1.5 & 90 & 57 & 45 & 120 & 0.01 & 0.01 & 0.01 & 0.01 & 0.98 \\
\F~\ref{fig:series_q2_low_e} & 10 & 10 & 5-9.1 & 0.91-5 & 1.0 & 2 & 20 & 0.1 & 0.4 & 1.5 & 90 & 0.01 & 45 & 0.01 & 0.01 & 0.01 & 0.01 & 0.01 & 0.98 \\
\F~\ref{fig:series_i2_low_e} & 10 & 10 & 5 & 5 & 1.0 & 2 & 20 & 0.1 & 0.4 & 1.5 & 90 & 0.01-70 & 45 & 0.01 & 0.01 & 0.01 & 0.01 & 0.01 & 0.98 \\
\bottomrule
\end{tabular}
\caption{Values of parameters used in the numerical integrations. Inclinations, arguments of periapsis, and longitudes of the ascending node are indicated with $i_i$, $\omega_i$ and $\Omega_i$, respectively (our reference frame is the $x,y$-plane, and the reference direction is the $x$-direction). The angles $\theta_1$ and $\theta_2$ are the true anomalies of orbits 1 and 2 (used only in the four-body integrations); $f_\theta$ indicates the fraction of the outer orbit true anomaly $\theta$ in the integrations compared to integrating from $t\rightarrow -\infty$ to $t\rightarrow \infty$ (see \eq~\ref{eq:tend}). Units of all angles ($i_1$, $i_2$, $\omega_1$, $\omega_2$, $\Omega_1$, $\Omega_2$, $\theta_1$ and $\theta_2$) are degrees. Units of masses are $\msun$ and distances are measured in $\au$ (note, however, that our system is scale free). }
\label{table:values}
\end{table}

\subsection{Changes of the outer orbit}
\label{sect:num:out}

\begin{figure}
\center
\includegraphics[scale = 0.4, trim = 0mm 10mm 0mm 10mm]{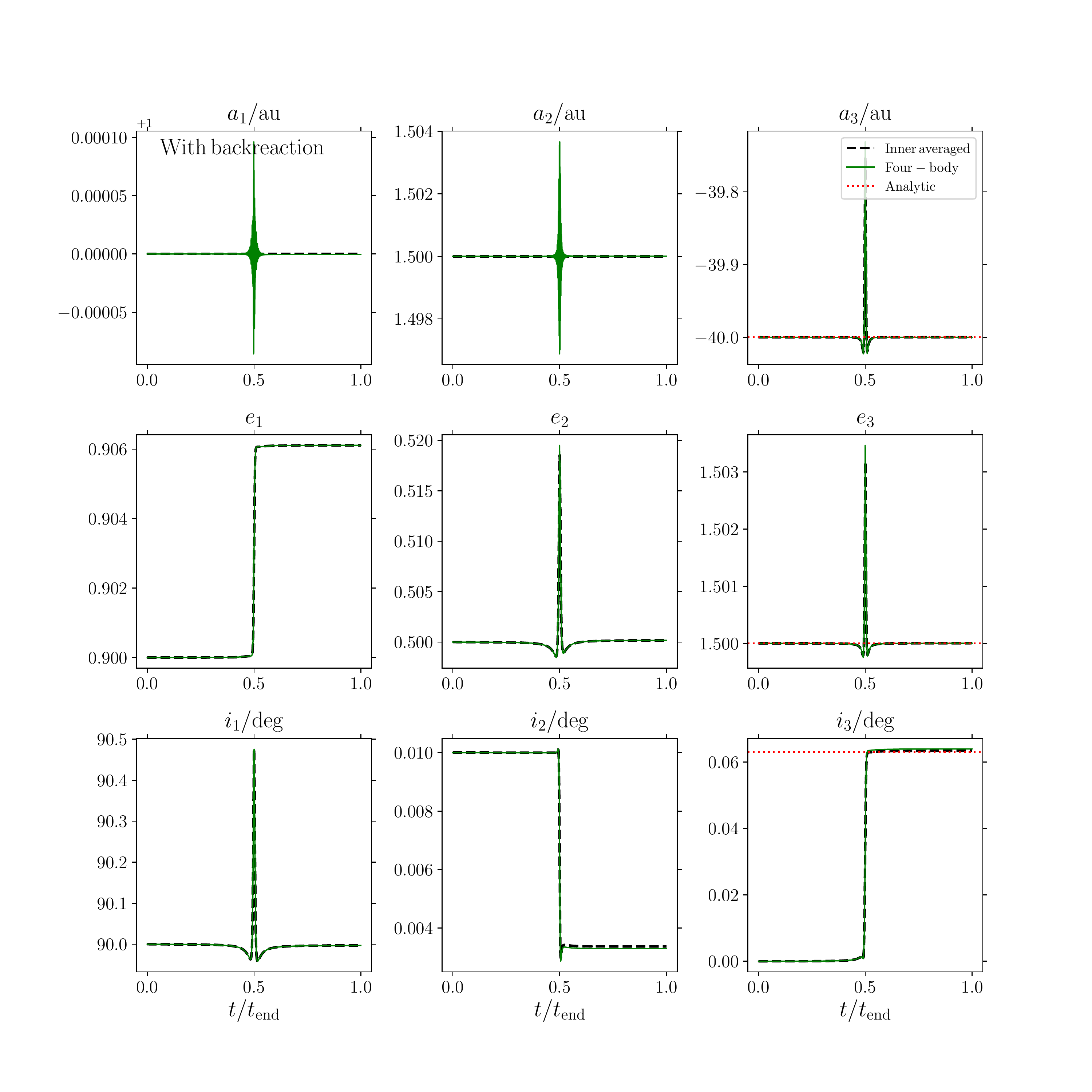}
\includegraphics[scale = 0.4, trim = 0mm 20mm 0mm 10mm]{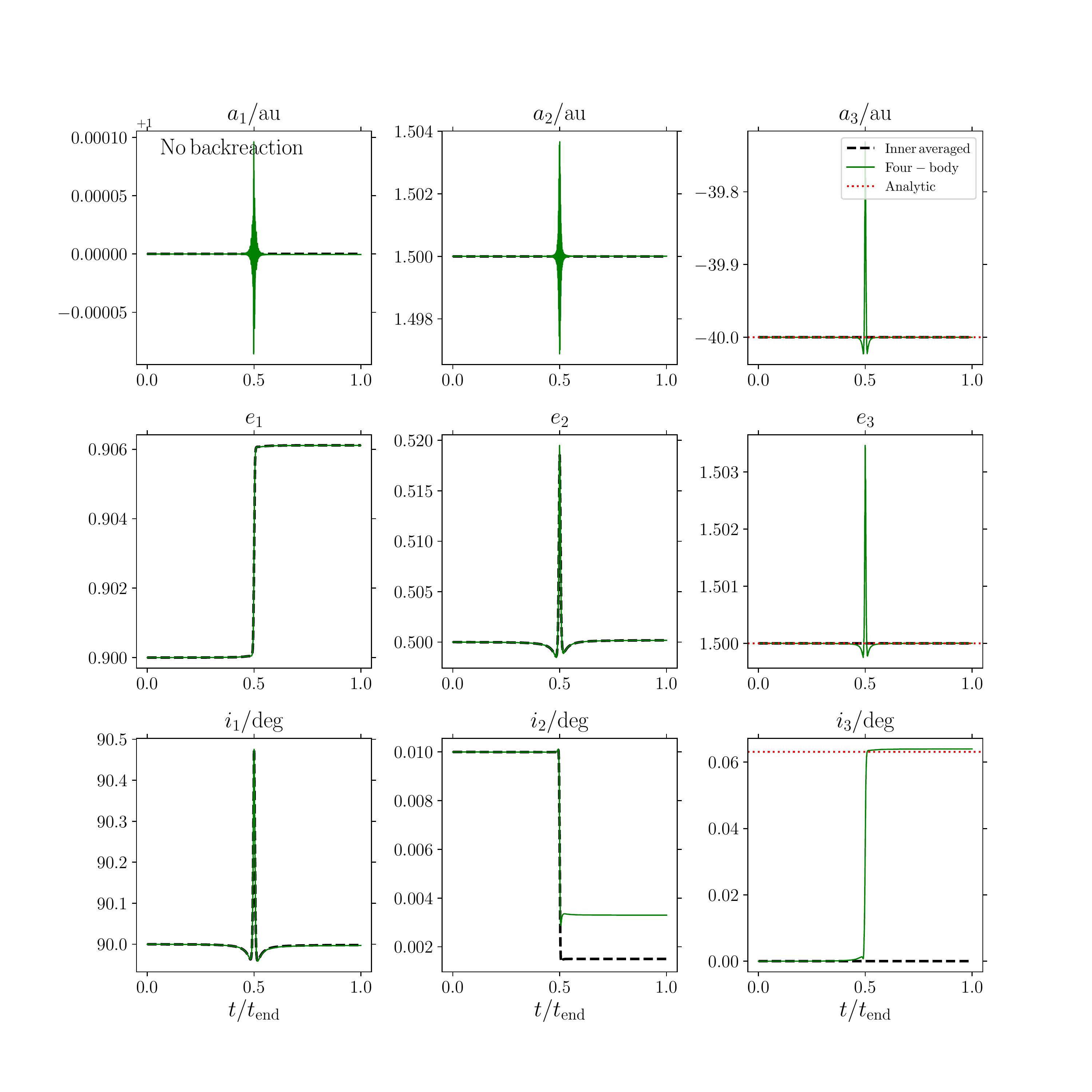}
\caption{ Evolution of the orbital elements of the three orbits as a function of time (normalised to the integration time, $t_\mathrm{end}$, see \eq~\ref{eq:tend}). See Table~\ref{table:values} for the initial conditions. The backreaction terms were included in the top nine panels, and excluded in the bottom nine panels. In each set of nine panels, the top row shows the semimajor axes, the middle row shows the eccentricities, and the bottom row shows the inclinations. Note that, initially, $i_3=0$ by the choice of the coordinate system. Solid green lines correspond to four-body integrations and black dashed lines to integrations averaged over the inner orbit (but not the outer orbit). In the third column, red dotted lines show analytic results for the net changes in the outer orbit (see \S~\ref{sect:an:an:out}). }
\label{fig:backreaction}
\end{figure}

As discussed in \S s~\ref{sect:an:ham:ex} and \ref{sect:an:an:out}, both binaries can affect the outer orbit and cause the latter to deviate from purely Keplerian motion. Consequently, this can affect the eccentricity and angular-momentum changes of the inner orbits, which we refer to as `backreaction'. In \F~\ref{fig:backreaction}, we show the time evolution of the orbital elements (semimajor axes, eccentricities and inclinations) of the three orbits. The top (bottom) nine panels correspond to the situation in which the backreaction terms to quadrupole order (see the expression for $\ddot{\ve{r}}_3$ in \eq~\ref{eq:eom}) were included (excluded). 

In each set of nine panels, the top row shows the semimajor axes. The four-body integrations (solid green lines) show tiny fluctuations in the semimajor axes near periapsis (note that in the top-left panel, $+1$ should be added in the $y$-axes). The inner-averaged integrations (black dashed lines) show no change in $a_1$ and $a_2$, as an immediate consequence of orbit averaging. When backreaction is included, the inner-averaged integrations give a fluctuation in $a_3$ near periapsis with no net change, and which agrees with the four-body integrations. The fact that the semimajor axes are conserved is expected for this system, which is well within the secular regime. 

The middle and bottom rows in each set of nine panels show the eccentricities and inclinations, respectively. Without backreaction, $e_3$ and $i_3$ in the inner-averaged integrations remain constant by construction, whereas the four-body integrations show that there is a net change in these quantities---the net change in $e_3$ is tiny, whereas it is more significant (but still very small) in $i_3$, with $\Delta i_3 \simeq 0.06^\circ$ in this case. With backreaction included, the inner-averaged integrations agree with the four-body integrations in terms of $e_3$ and $i_3$. Also, the analytic prediction for $\Delta i_3$ agrees with the numerical results. 

Moreover, in terms of the inner orbit eccentricities and inclinations from the inner-averaged integrations and comparing the top and bottom set of nine panels, it is clear that the backreaction terms have no appreciable effect (the only noticeable effect is a slight different in $i_2$ of $\simeq 0.002^\circ$, as shown in the bottom-middle panel of the low set of nine panels).

\subsection{The impact of the `binarity' of the companion}
\label{sect:num:bin}

\begin{figure}
\center
\includegraphics[scale = 0.65, trim = 0mm 10mm 0mm 0mm]{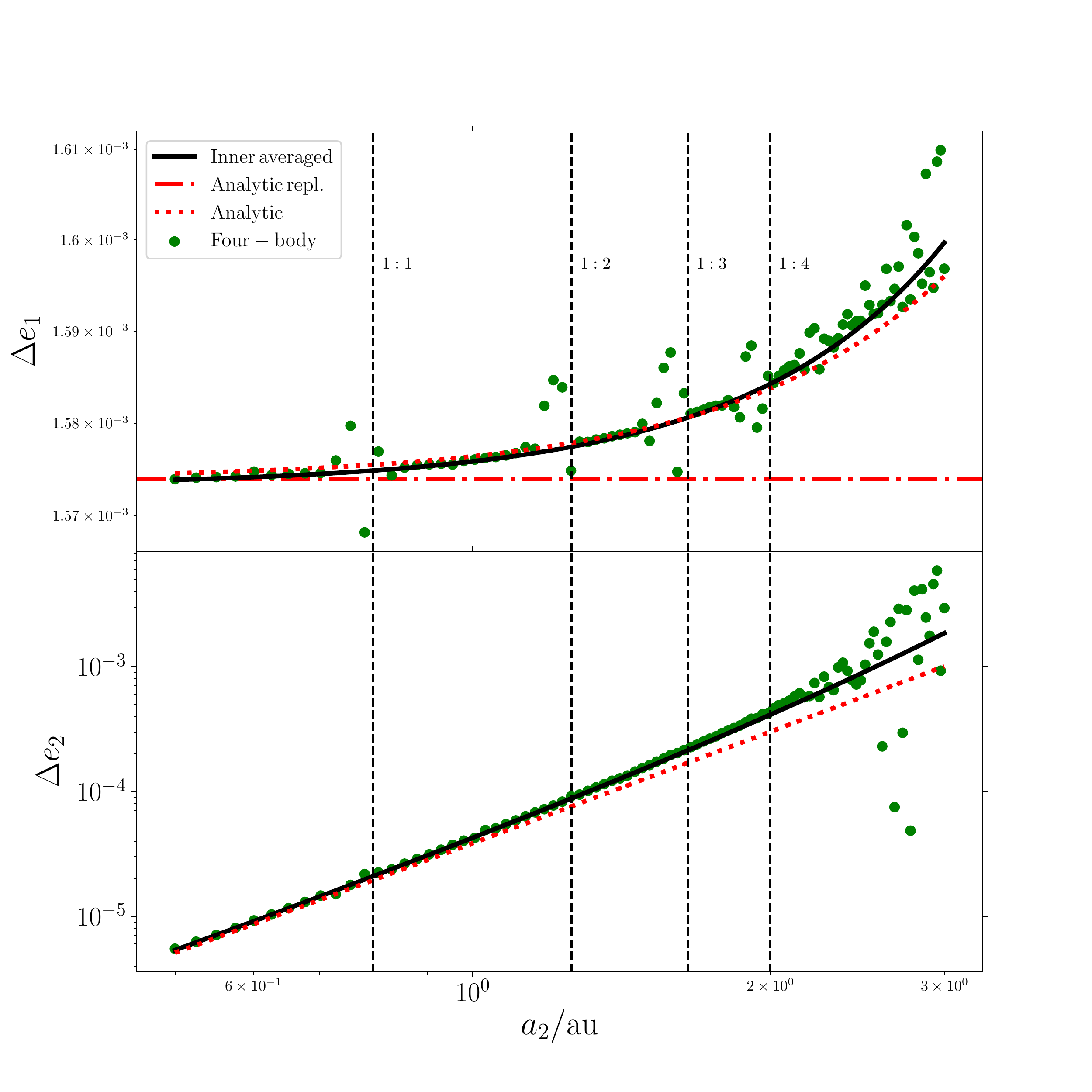}
\caption{ Scalar eccentricity changes in orbits 1 (top panel) and 2 (bottom panel) as a function of $a_2$. See Table~\ref{table:values} for the initial conditions. Green dots correspond to four-body integrations, solid black lines to inner-averaged integrations (`Inner averaged' in the legend), and red lines to analytic expressions (see \S~\ref{sect:an:an:in}). For the red dotted lines (`Analytic' in the legend), the hexadecupole-order cross term is included, whereas it is not for the horizontal red dot-dashed line (`Analytic repl.', i.e., `Analytic replaced' in the legend). In other words, orbit 2 is considered to be a point mass in the `Analytic repl.' horizontal red dot-dashed lines.}
\label{fig:series_a2_low_e}
\end{figure}

\begin{figure}
\center
\includegraphics[scale = 0.65, trim = 0mm 10mm 0mm 0mm]{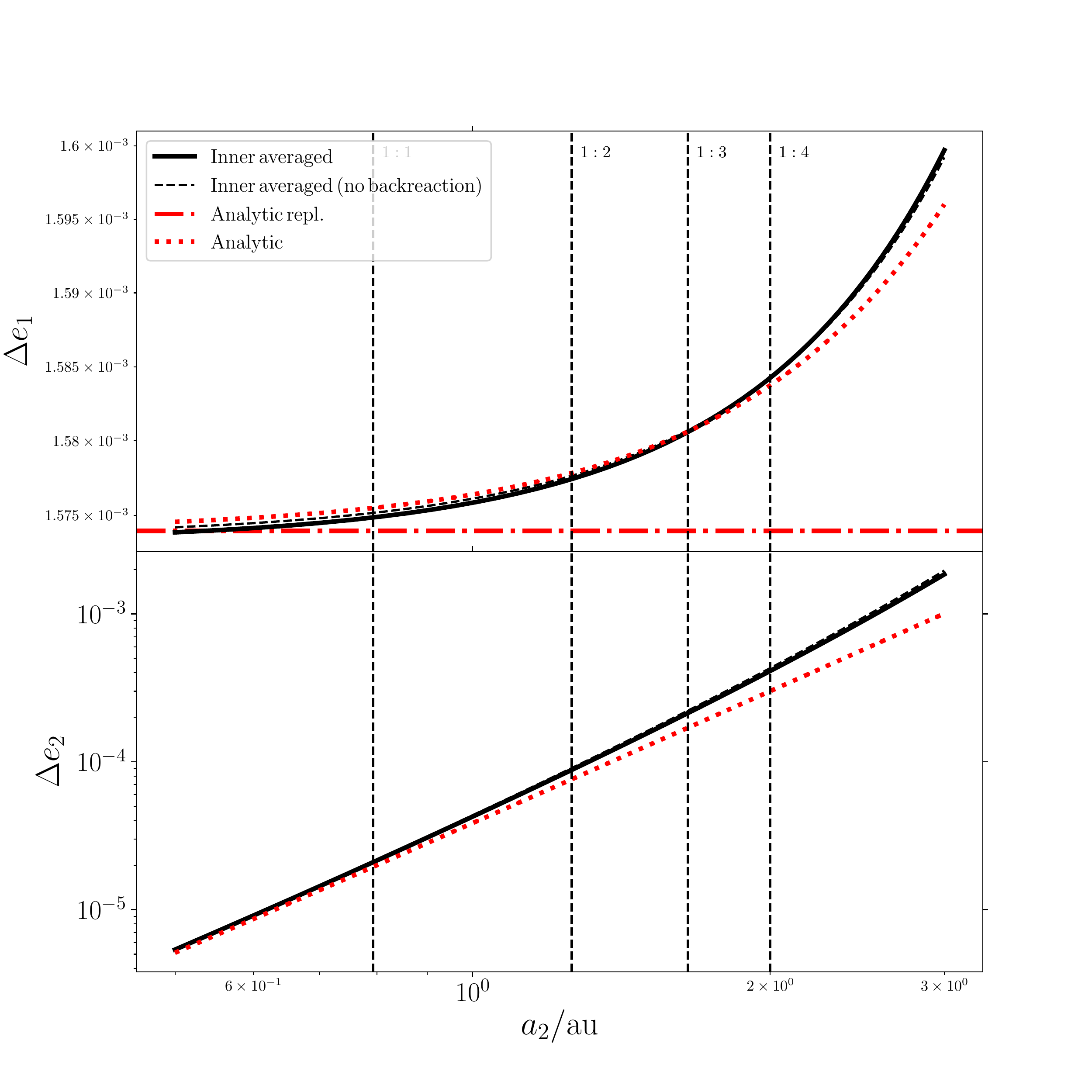}
\caption{ Similar to \F~\ref{fig:series_a2_low_e}, but here comparing two cases of inner-averaged numerical integrations: with backreaction on the outer orbit (black solid lines), and without (black dashed lines). The analytic curves (red lines) are the same as in \F~\ref{fig:series_a2_low_e}. }
\label{fig:series_a2_low_e_compare_back}
\end{figure}

\begin{figure}
\center
\includegraphics[scale = 0.65, trim = 0mm 10mm 0mm 0mm]{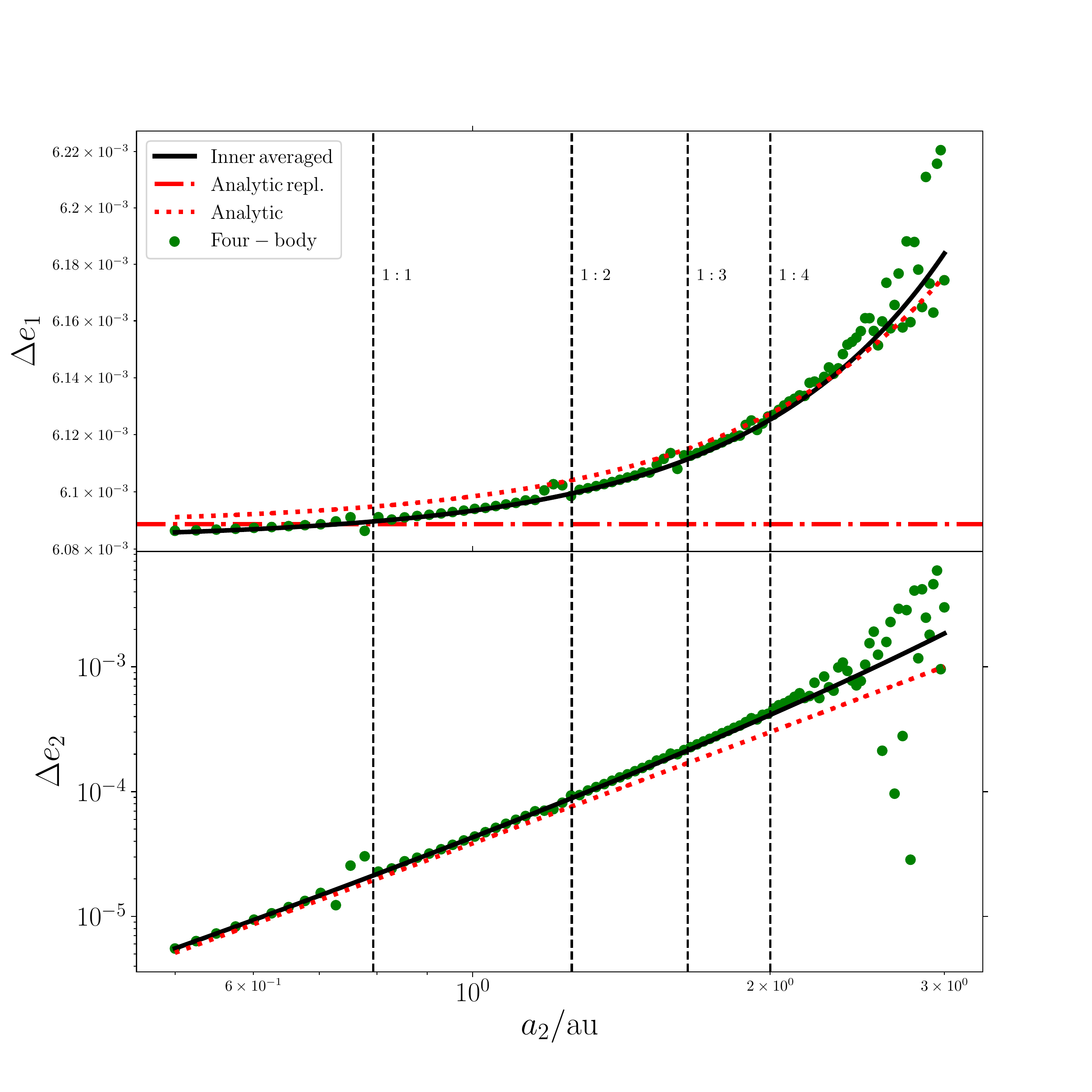}
\caption{ Similar to \F~\ref{fig:series_a2_low_e}, but here with higher initial $e_1$ (see Table~\ref{table:values}). }
\label{fig:series_a2_high_e}
\end{figure}

\begin{figure}
\center
\includegraphics[scale = 0.65, trim = 0mm 10mm 0mm 0mm]{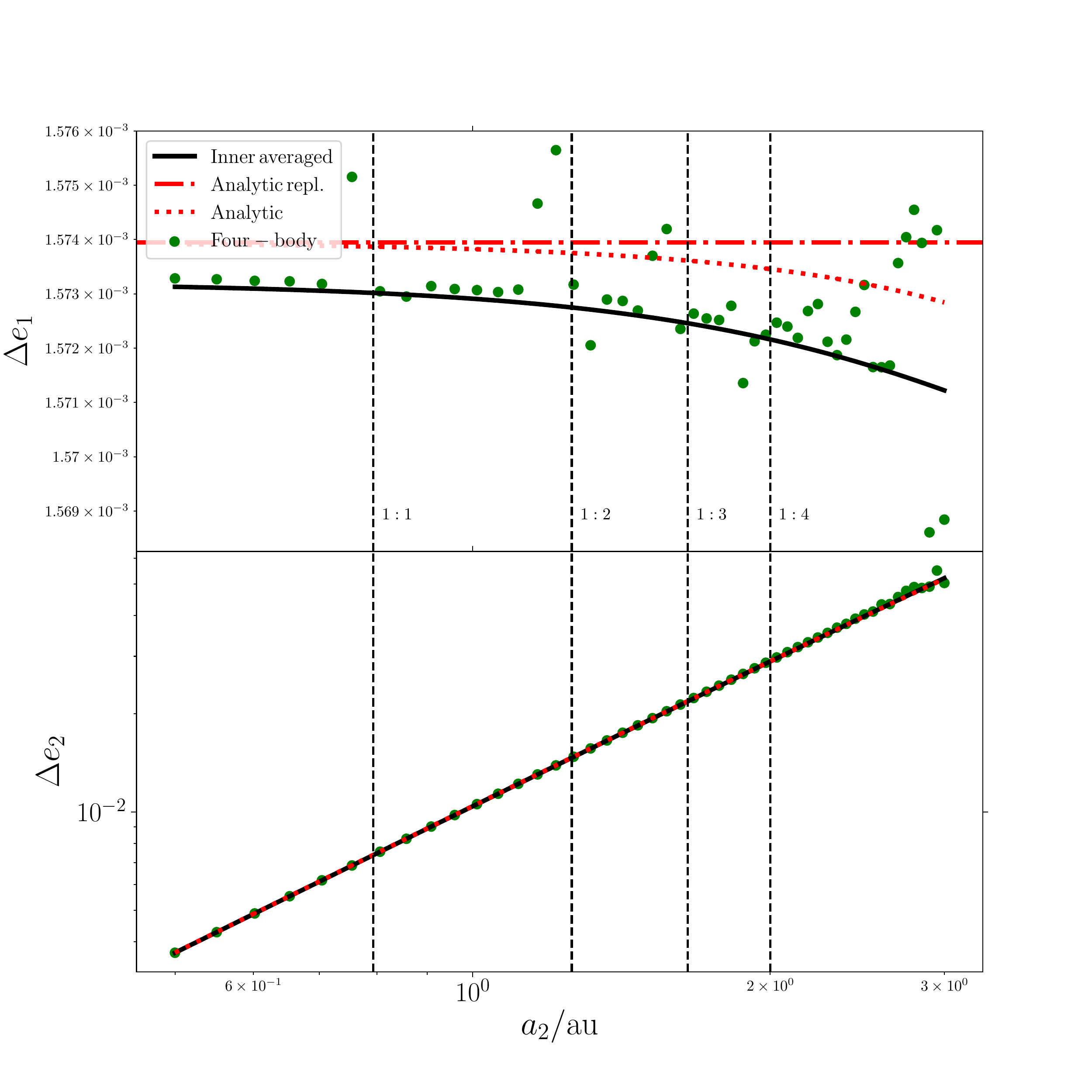}
\caption{ Similar to \F~\ref{fig:series_a2_low_e}, but here with different initial $\omega_2$ and $i_2$ (see Table~\ref{table:values}). }
\label{fig:series_a2_angle}
\end{figure}

\begin{figure}
\center
\includegraphics[scale = 0.65, trim = 0mm 10mm 0mm 0mm]{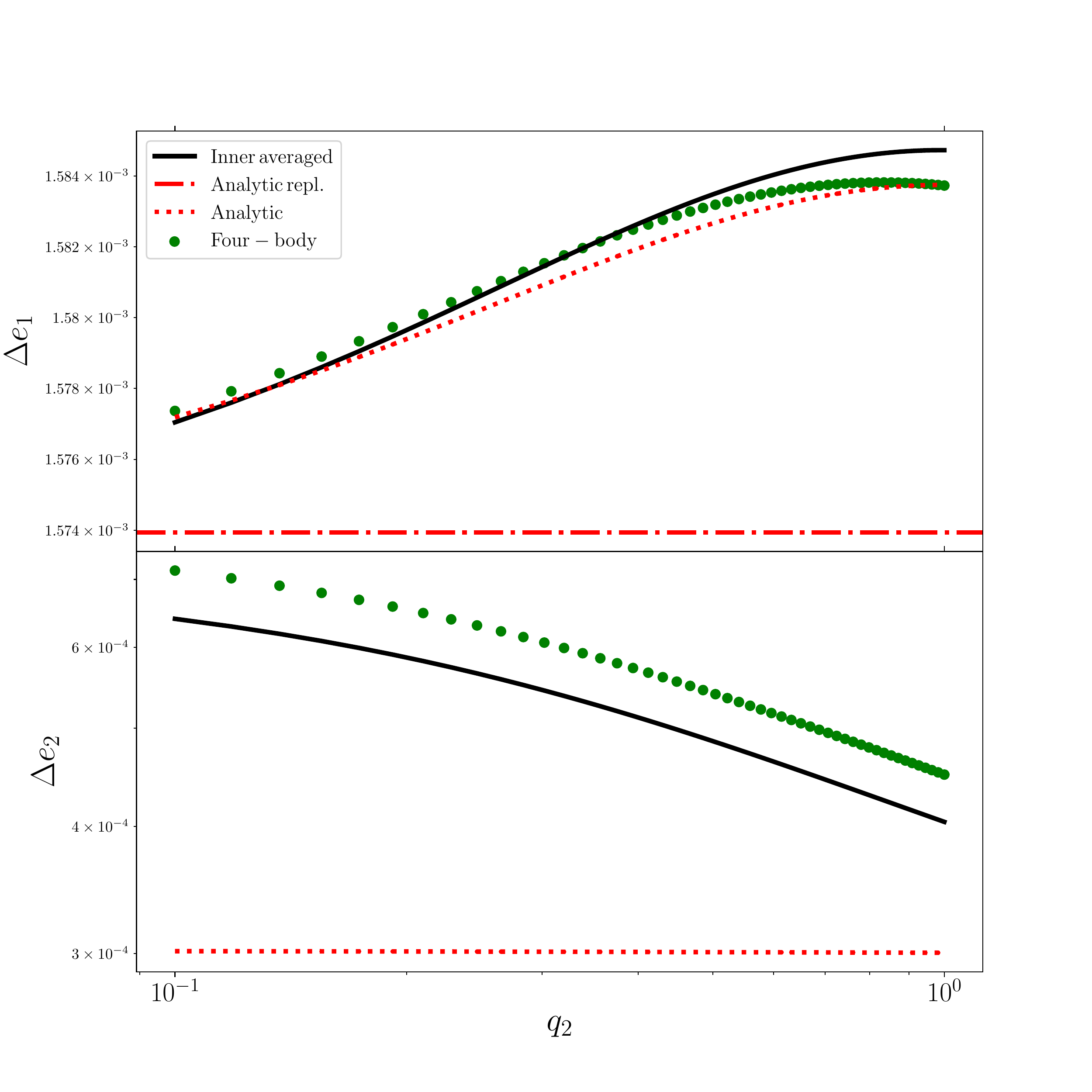}
\caption{ Similar to \F~\ref{fig:series_a2_low_e}, but here with fixed $a_2$ and varying $m_3$ and $m_4$, keeping $M_2\equiv m_3+m_4 = 10 \, \msun$ fixed and plotting the eccentricity changes as a function of $q_2 \equiv m_4/m_3$ (see also Table~\ref{table:values}). }
\label{fig:series_q2_low_e}
\end{figure}

\begin{figure}
\center
\includegraphics[scale = 0.65, trim = 0mm 10mm 0mm 0mm]{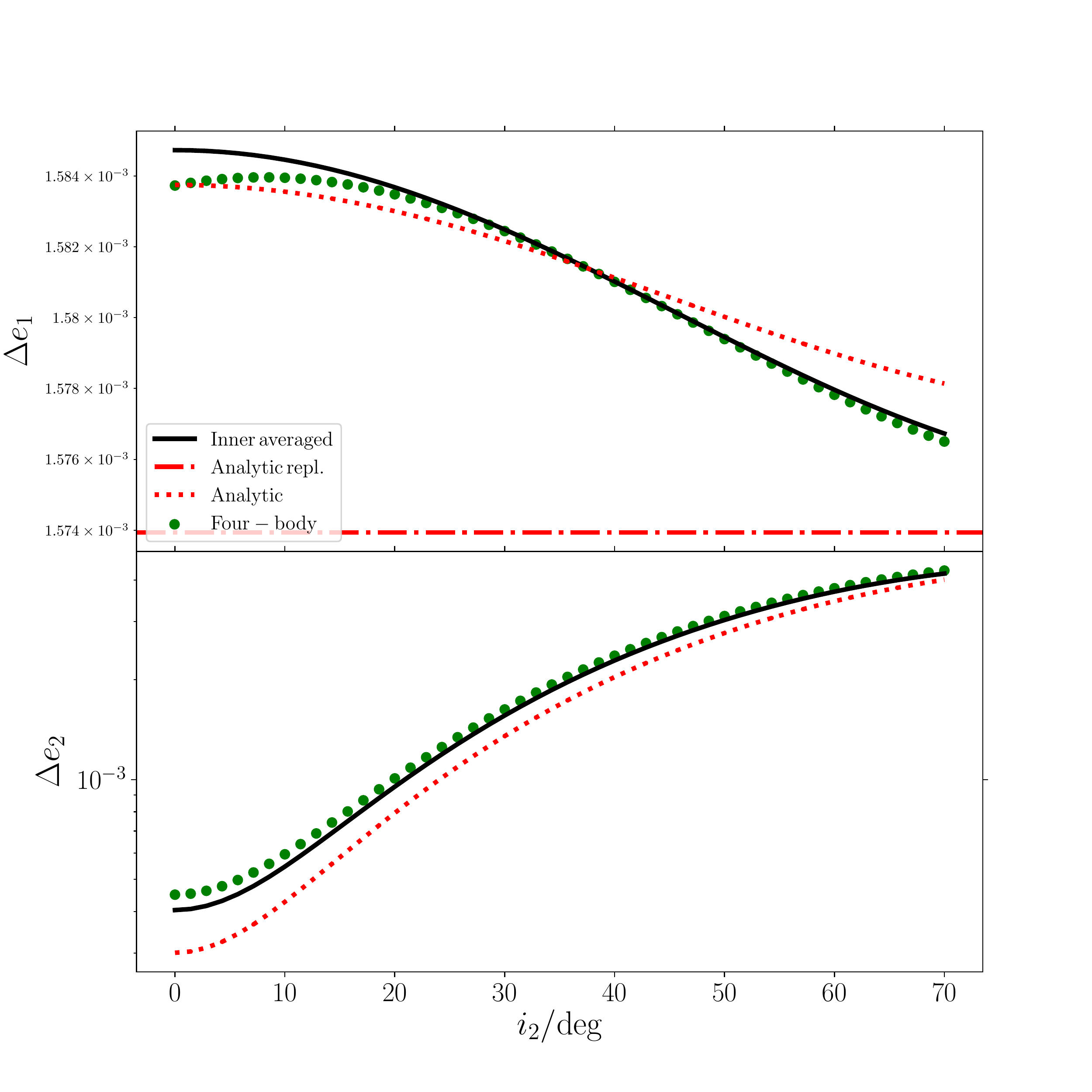}
\caption{ Similar to \F~\ref{fig:series_a2_low_e}, but here with fixed $a_2$ and varying $i_2$ (see also Table~\ref{table:values}). }
\label{fig:series_i2_low_e}
\end{figure}

Here, we carry out several series of integrations to investigate the effect that the `binarity' of orbit 2 has on the eccentricity change of orbit 1. All initial conditions can be found in Table~\ref{table:values}. 

In \F~\ref{fig:series_a2_low_e}, we vary $a_2$, keeping all other parameters fixed. Evidently, as $a_2 \rightarrow 0$, we recover the limit of an encounter of a binary with a single point mass. First of all, note that the eccentricity changes in orbit 1 are very weakly dependent on $a_2$: varying $a_2$ between 0.5 and 3 $\au$ affects $\Delta e_1$ by only $\sim 10^{-5}$. Even for the largest value of $a_2$ considered, it is still a good approximation to consider orbit 2 as a point mass in the computation of $\Delta e_1$ (see the red horizontal dot-dashed line in \F~\ref{fig:series_a2_low_e}, which shows the corresponding analytic value assuming binary 2 is a point mass). Note that if $a_2$ is much larger than $3\,\au$, the system would no longer be in the secular regime (cf. \eq~\ref{eq:Ri}). The eccentricity of orbit 2 changes much more appreciably and according to a power law, which is expected given that $a_2$ is varied in this series of integrations (e.g., \citealt{1996MNRAS.282.1064H}). 

The inner-averaged integrations generally agree with the four-body results, i.e., $\Delta e_1$ increases with increasing $a_2$ with our choice of initial conditions. Some deviations are apparent at specific values of $a_2$, as well as for larger values of $a_2$. The latter can be explained by the fact that $\epssa{2}$ is approaching unity as $a_2$ increases, with $\epssa{2} \simeq 0.017$ if $a_2=3\,\au$, i.e., the system gradually becomes less secular. The discrepancies at smaller values of $a_2$ are likely due to mean-motion resonances (MMRs) between the two inner orbits. This is suggested by their occurring locations in $a_2$, which correspond to various MMRs and which are indicated with vertical black dashed lines. Specifically, we show the $1:\alpha$ resonances between orbits 1 and 2, where $\alpha\in \{1,2,3,4\}$; setting $P_1 = \alpha P_2$, where $P_i$ denotes orbital period and $\alpha$ is a dimensionless factor, implies
\begin{align}
a_2 = a_1 \left [\alpha^2 (M_2/M_1)  \right ]^{1/3}. 
\end{align}

The red dotted lines in \F~\ref{fig:series_a2_low_e} show the analytic results from \S~\ref{sect:an:an:in} with the inclusion of terms up to and including hexadecupole order (and including the cross term), as well as the quadrupole-order term that is second order in $\epssa{i}$ (see \citealt{2019MNRAS.487.5630H}). Overall, these analytic expressions agree with the numerical results, although some deviation can be seen, especially for larger $a_2$. This can be attributed to the fact that the analytic expressions do not fully take into account the changing $e_i$ during the encounter (only to second order in $\epssa{i}$, and at quadrupole expansion order). Although it is possible in principle to derive more accurate expressions, they are excessively long and so are not practical (see, e.g., table 1 of \citealt{2019MNRAS.488.5192H}). 

In \F~\ref{fig:series_a2_low_e_compare_back}, we consider the same series as in \F~\ref{fig:series_a2_low_e}, but include only inner-averaged numerical integrations, and compare the cases including backreaction on the outer orbit (black solid lines), and without (black dashed lines). As shown, there are only very small differences between the two cases, again illustrating that the backreaction of the inner two orbits on the outer orbit can be neglected. 

We show a similar figure to \F~\ref{fig:series_a2_low_e} in \F~\ref{fig:series_a2_high_e}, but now with a higher initial value of $e_1$. The eccentricity changes are now slightly larger, and the relative importance of MMRs appears to be lower. In \F~\ref{fig:series_a2_angle}, we choose different values of $\omega_2$ and $i_2$. With this different choice of relative orbital orientations, the changes in $\Delta e_1$ with increasing $a_2$ are even smaller, on the order of $\sim 10^{-6}$. The different relative orientation between the orbits in this example leads to a decrease in $\Delta e_1$ with increasing $a_2$, instead of increasing in Figs~\ref{fig:series_a2_low_e} and \ref{fig:series_a2_high_e}. Also, MMRs appear to have a larger impact on $\Delta e_1$. 

In \F~\ref{fig:series_q2_low_e}, we fix $a_2$ but vary $q_2 \equiv m_4/m_3$ instead, keeping the total mass of binary 2, $M_2 =m_3+m_4$, fixed to $M_2 = 10\,\msun$. The point-mass limit is approached as $q_2\rightarrow 0$, and $\Delta e_1$ indeed approaches the point-mass value (red horizontal dot-dashed line) as $q_2$ decreases. The dependence on $q_2$, like $a_2$, is very weak, with changes in $\Delta e_1$ on the order of $10^{-5}$. The analytic results including the hexadecupole-order cross term (red dotted lines) agree with the numerical results (both four-body and inner-averaged), except for $\Delta e_2$. This may be related to the omission of higher-order terms in $\epssa{i}$. In addition, the four-body integrations do not agree well with the inner-averaged integrations with respect to $\Delta e_2$, which may be due to a breakdown of the inner-averaging approximation. 

Lastly, in \F~\ref{fig:series_i2_low_e}, we consider the dependence on relative orientation by varying $i_2$ and fixing the other parameters. The changes in $e_1$ are again very small, and $\Delta e_1$ decreases with increasing $i_2$. The analytic expressions agree reasonably with the numerical results.

\section{Discussion}
\label{sect:discussion}
\subsection{Importance of the cross term}

\begin{figure}
\center
\includegraphics[scale = 0.38, trim = 0mm 0mm 0mm 0mm]{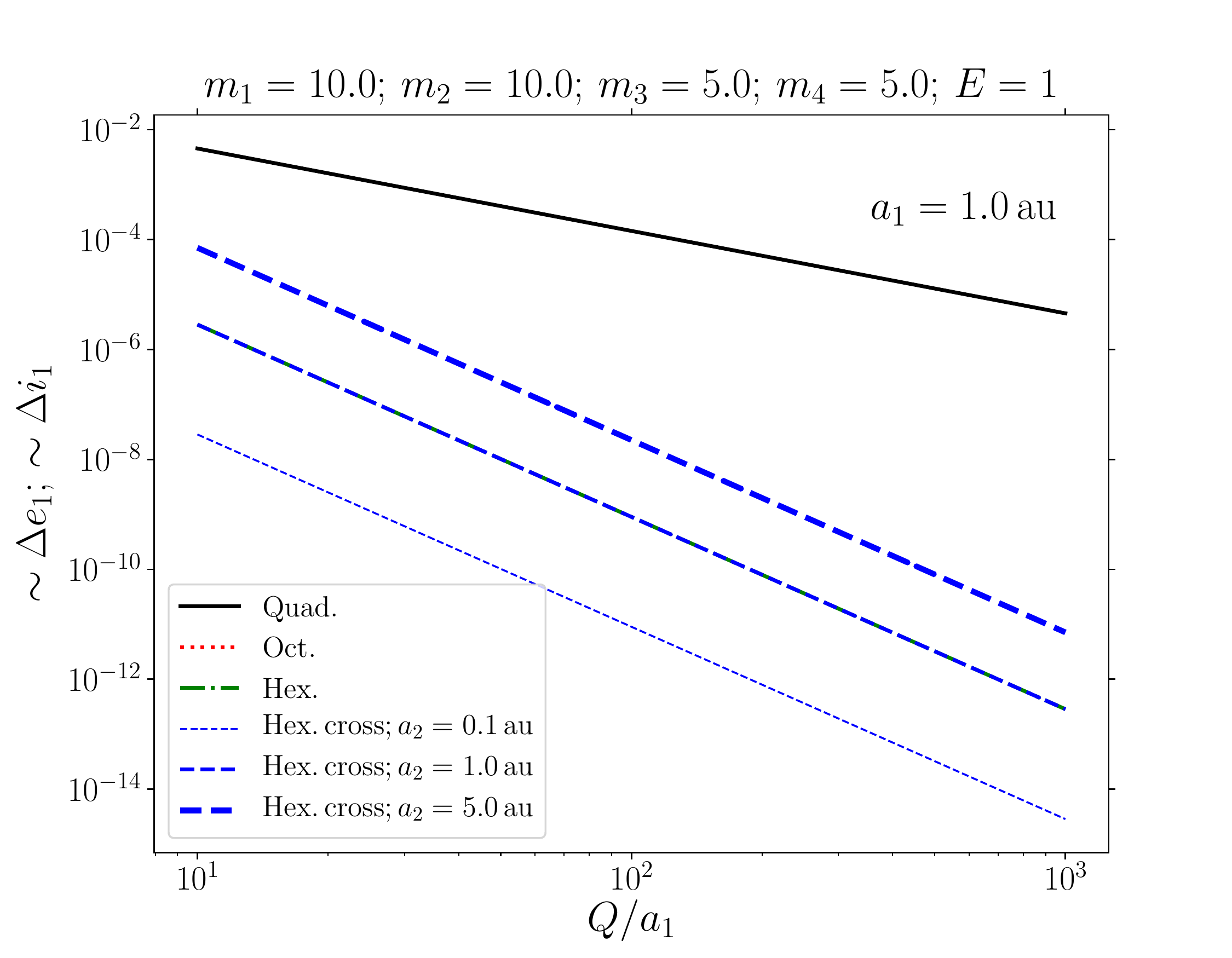}
\includegraphics[scale = 0.38, trim = 0mm 0mm 0mm 0mm]{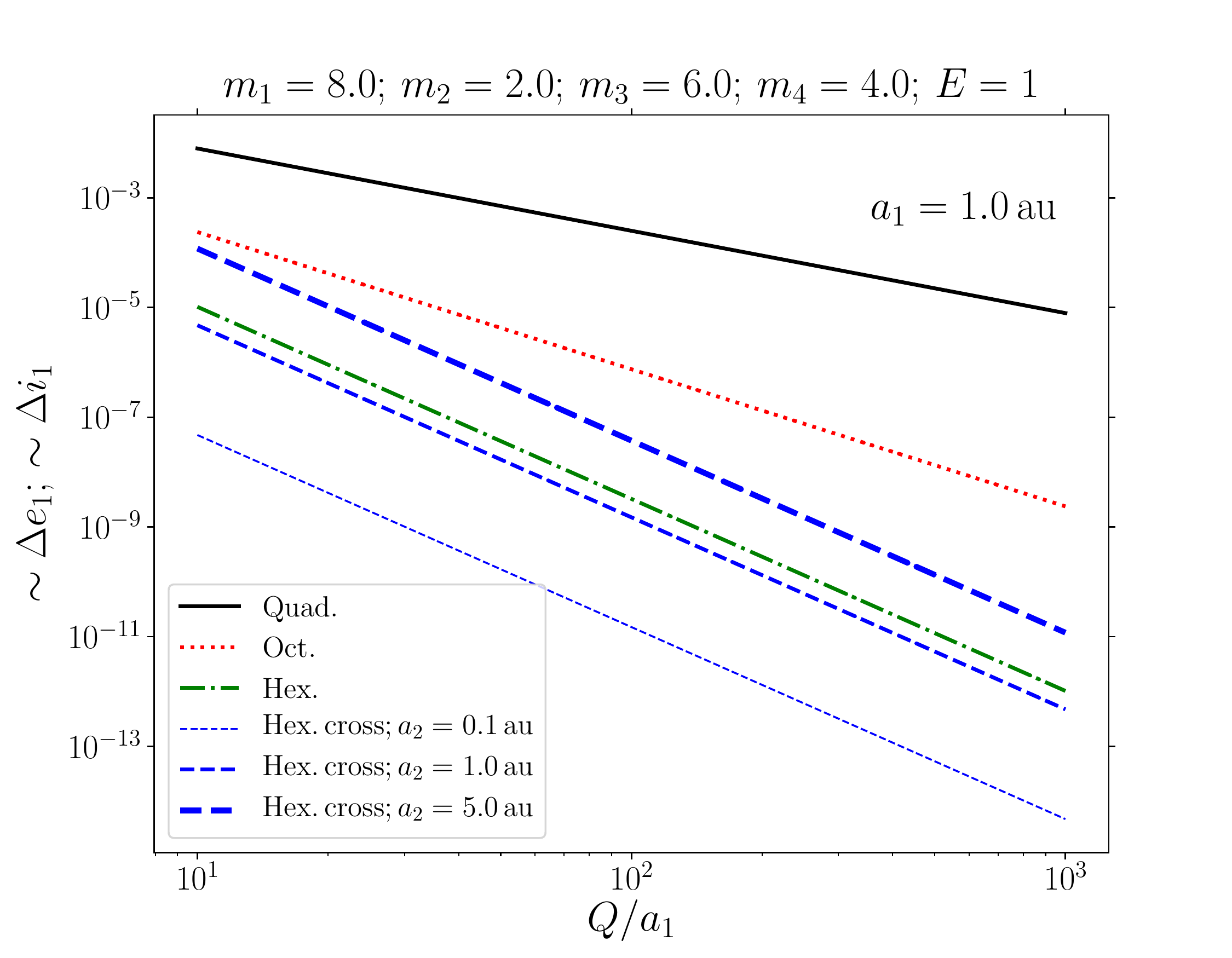}
\includegraphics[scale = 0.38, trim = 0mm 0mm 0mm 0mm]{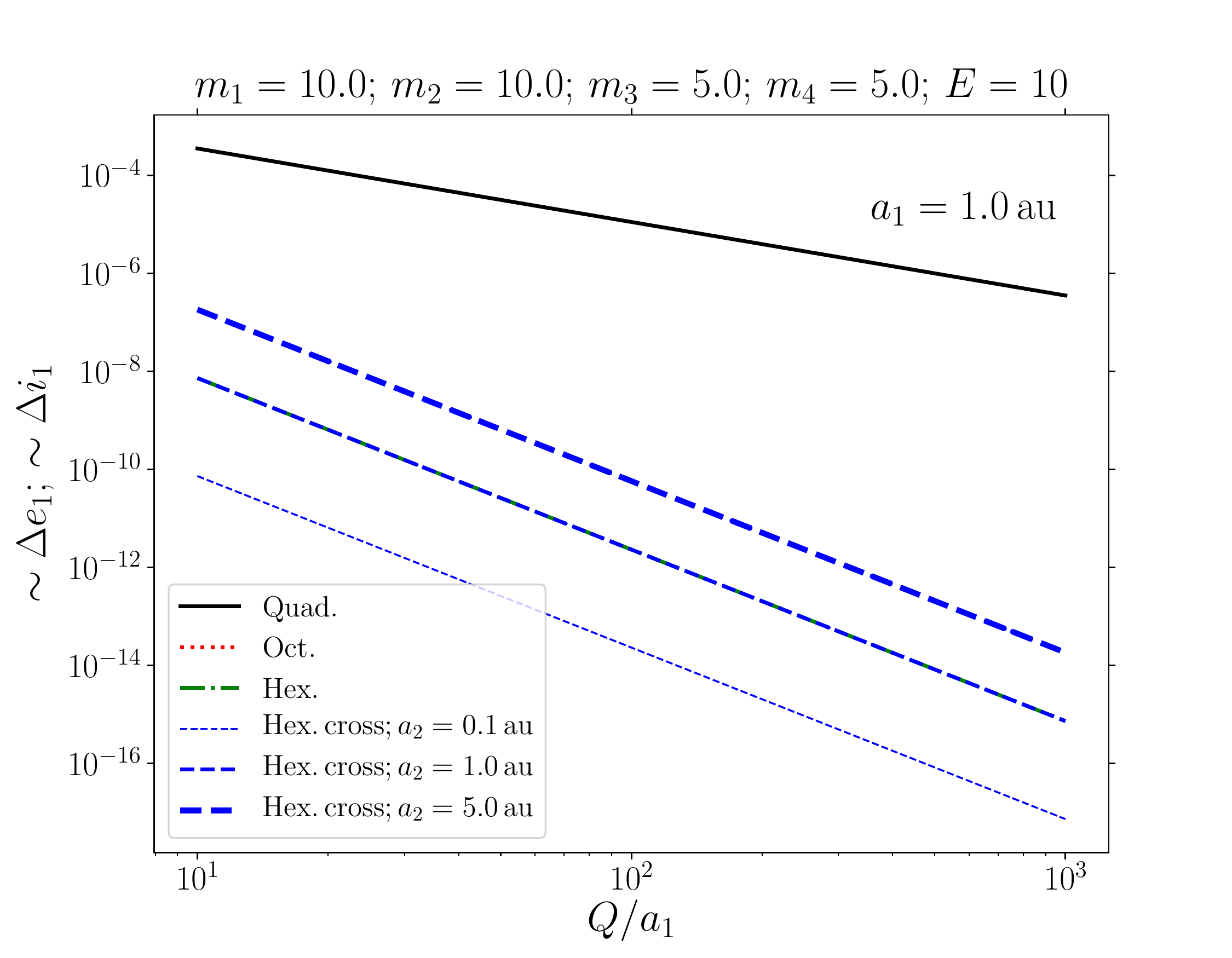}
\includegraphics[scale = 0.38, trim = 0mm 0mm 0mm 0mm]{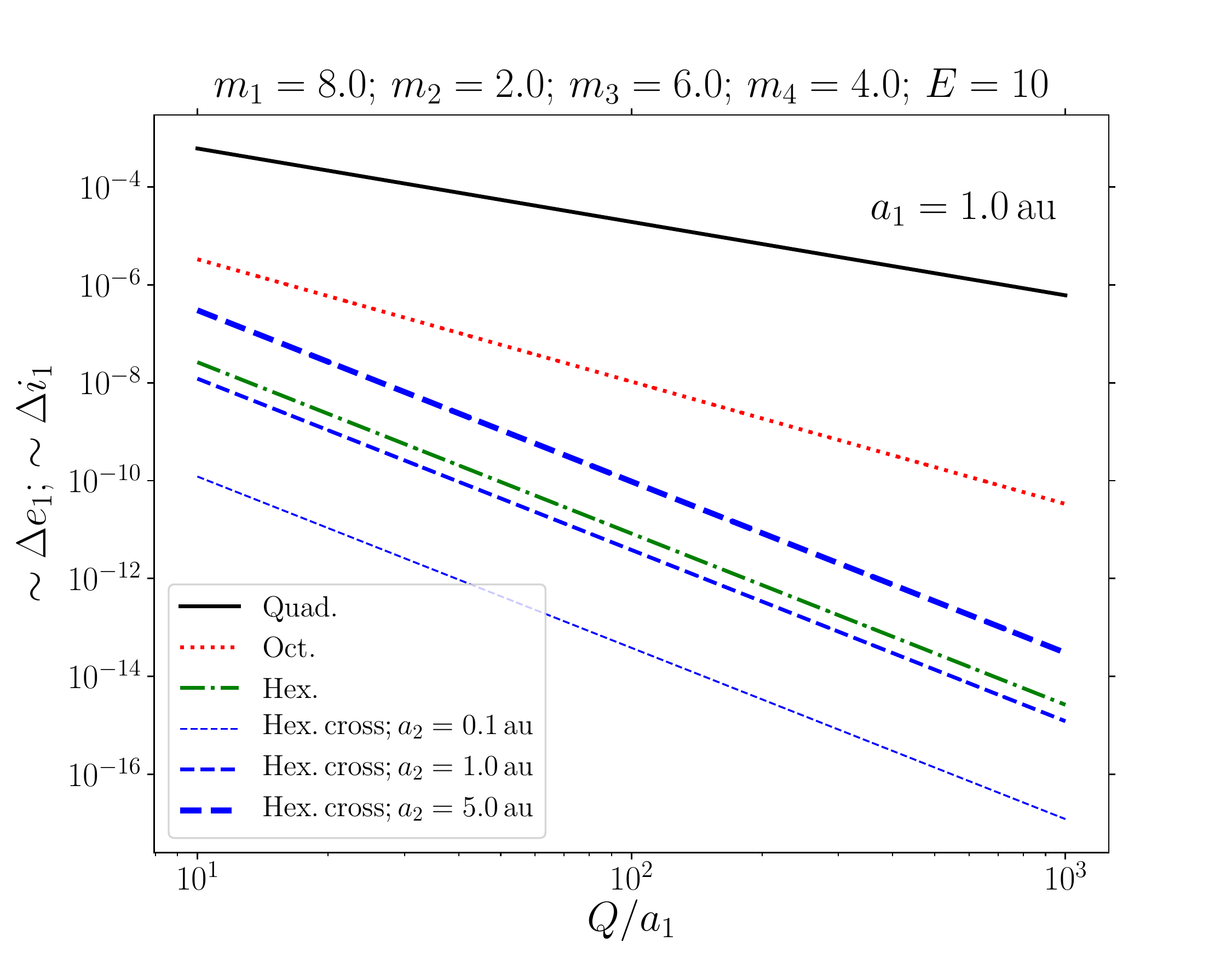}
\caption{ Changes of the eccentricity and inclination of orbit 1 plotted as a function of $\qper/a_1$ for a fixed $a_1=1\,\au$ and different $a_2$ (the latter being either 0.1, 1.0, or 5.0 $\au$). We show the contributions from various expansion orders in the Hamiltonian to $\Delta e_1$ and $\Delta i_1$, which we roughly estimate (within approximately an order of magnitude) using \eq~(\ref{eq:Delta_e1_est}). Each panel corresponds to a certain choice of the masses $m_i$ and the outer orbit eccentricity $\eper$, indicated in the top. Note: in the left-hand column, the `Oct.' lines are zero and therefore not shown, since $m_1=m_2$. Also, the `Hex.' lines coincide exactly with the `Hex. cross' lines at $a_2=1\,\au$. }
\label{fig:overview}
\end{figure}

As shown in the above sections, in the expansion of the Hamiltonian of the system, the hexadecupole order is the lowest expansion order at which a term appears that explicitly depends on all three orbits simultaneously (the inner two bound orbits and the outer unbound orbit). This `cross term' gives rise to the largest changes of the secular changes of one binary due to the `binarity' of the other binary. Given that the cross term appears at a high expansion order, the `binarity' effect of the companion binary is typically small and, in most cases, it is well justified to simply apply the known expressions for the secular changes for binary-single interactions \citep{1996MNRAS.282.1064H,2009ApJ...697..458S,2018MNRAS.476.4139H,2019ApJ...872..165G,2019MNRAS.487.5630H,2019MNRAS.488.5192H} with the `third body' mass replaced with the total mass of the companion binary. 

Nonetheless, it is informative to explore more generally the importance of the hexadecupole-order cross term in relation to the other terms of interest, i.e., the pairwise quadrupole, octupole, and hexadecupole-order terms. In \F~\ref{fig:overview}, we estimate (i.e., within approximately an order of magnitude) the changes in eccentricity and inclination of orbit 1, plotting their rough approximations as a function of $\qper/a_1$ for fixed $a_1$, and various values of $a_2$. Here, we estimate the eccentricity/inclination changes based on \eq~(\ref{eq:Delta_e_i_FO}), ignoring the complex dimensionless functions of $\ve{e}_i$, $\ve{\j}_i$ and $\eper$ and any terms $\mathcal{O} \left (\epssa{i}^2\right)$. Specifically, we set
\begin{subequations}
\begin{align}
\label{eq:Delta_e1_est}
\Delta e_{1,\mathrm{quad}} &\sim \epssa{1}; \\
\Delta e_{1\mathrm{oct}} &\sim \epssa{1} \epsoct{1}; \\
\Delta e_{1,\mathrm{hex}} & \sim \epssa{1} \epshex{1}; \\
\Delta e_{1,\mathrm{hex,cross}} &\sim \epssa{1} \epshexcross{1},
\end{align}
\end{subequations}
and similarly for $\Delta i_1$. 

\F~\ref{fig:overview} shows that the cross term is small, and can be neglected in most practical situations. It is possible that the cross term exceeds the contribution from the pairwise hexadecupole-order term, but only in situations with large $a_2$ ($a_2=5\,\au$ in our examples), in which case the system is barely in the secular limit (note that the smallest $\qper$ in \F~\ref{fig:overview} is $\qper = 10\,\au$, which is only twice as large).

\subsection{Limitations of the analytic expressions and the inner-averaged approach}
\label{sect:discussion:lim}
In \S~\ref{sect:an:an:in}, we derived analytic expressions for the eccentricity changes taking into account the hexadecupole-order cross term, which is the lowest-order term that leads to a direct coupling between the inner two orbits. These expressions agree reasonably with numerical integrations, both four-body integrations and inner-averaged integrations, although the agreement is by no means perfect. Any deviation between the inner-averaged integrations and the analytic expressions arises from the fact that we assumed in \S~\ref{sect:an:an:in} that all three orbits are static during the encounter. This approximation can break down, especially when the initial eccentricities are already large (making the inner orbits more susceptible to large secular changes). 

Corrections to counter the breakdown of this approximation could be derived to second (and higher) order in $\epssa{i}$, as has been done in \citet{2019MNRAS.487.5630H}. When comparing to numerical results in \S~\ref{sect:num}, we also included second-order terms in $\epssa{i}$, but only to the quadrupole order. Similar expressions to higher orders in $\epssa{i}$ give rise to excessively long expressions (see \citealt{2019MNRAS.488.5192H}), which severely reduces their practical usefulness. Moreover, contributions from the second-order terms in $\epssa{i}$ at higher expansion orders (octupole, hexadecupole, etc.) become increasingly small. Here, we therefore did not derive new expressions for the eccentricity changes taking into account nonstatic orbits during the encounter for high expansion orders (in particular, for the hexadecupole-order cross term). 

In addition, we found discrepancies between the four-body integrations and the inner-averaged integrations (on which the fully analytic expressions are based). This is reflected in \F~\ref{fig:series_a2_low_e} and further, where the inner-averaged (black solid lines) and fully-averaged (red dotted lines) show disagreement with the four-body integrations near the mean motion resonance locations, as well as for large $a_2$, when the averaging approximation breaks down because $a_2$ is becoming too large. These discrepancies can be attributed to a breakdown of averaging in the inner orbits. Averaging corrections to the inner orbits as well (see, e.g., \citealt{2019MNRAS.490.4756L}) are beyond the scope of this paper.

\subsection{Implications for larger-$N$ scattering in the secular limit}
\label{sect:discussion:larger}
We briefly discuss further implications of the main result of this paper, i.e., that, in the secular limit, a binary perturbed by another distant binary is not significantly affected by the quadrupole moment of the companion binary, and that the orbital changes can simply be obtained by applying known results for binary-single encounters and replacing the third body's mass with the total companion binary mass. With this result in mind, it is clear that an extension to encounters with higher-multiplicity systems in the secular limit can easily be made: for a binary encountering an arbitrary hierarchical system composed of nested orbits one can, to first approximation, apply the secular expressions for binary-single encounters (possibly including higher-order terms in $\epssa{1}$) replacing the `third body' mass with the total mass of the encountering hierarchical system. This also implies that any internal evolution of the encountering system does not play any major role, no matter its own evolution timescale in relation to the encounter timescale. For example, a binary encountering a triple results in approximately the same secular effects on the binary compared to the case of a binary encountering a single object with the same mass as the triple.

\section{Conclusions}
\label{sect:conclusions}
We studied the dynamical evolution of two binaries approaching each other on unbound orbits. We focused on the `secular' regime, in which the binaries approach each other with a sufficiently large periapsis distance such that the semimajor axes of the two bound orbits do not change appreciably after the encounter, but eccentricity and angular-momentum changes are possible. We carried out numerical integrations, as well as derived analytic results. Our main conclusions are given below.

\medskip \noindent 1. The Hamiltonian, expanded in the small ratios $x_1=r_1/r_3\ll 1$ and $x_2=r_2/r_3 \ll 1$, where $r_1$ and $r_2$ are the relative separations of the inner two bound binaries and $r_3$ is the separation of the `outer' unbound orbit, consists of pairwise terms at the quadrupole and octupole orders ($x_i=2$ and $x_i=3$, respectively, for $i\in\{1,2\}$). Only at the hexadecupole order ($x_i=4$) there appears a term, the hexadecupole-order cross term, that explicitly depends on the separations of all three orbits. This implies that any effect of the `binarity' of orbit 2 on orbit 1 (i.e., its quadrupole moment) is only exhibited through 1) a backreaction of the outer orbit, $\ve{r}_3$, and 2) high-order expansion terms, starting at the hexadecupole order. We explicitly derived the expanded Hamiltonian (up to and including hexadecupole order) and averaged over the inner two orbits (\S~\ref{sect:an:ham:oa}), as well as the corresponding equations of motion (\S~\ref{sect:an:eom}). 

\medskip \noindent 2. We derived approximate analytic expressions for the eccentricity and inclination changes of the outer orbit due to the backreaction of orbits 1 and 2 (\S~\ref{sect:an:an:out}). These expressions show that the backreaction effects are very small, which we confirmed with numerical integrations (\S~\ref{sect:num:out}).  

\medskip \noindent 3. We also derived approximate analytic expressions for the secular effects on the inner orbits taking into account the hexadecupole-order cross term. In particular, the quadrupole moment of the companion orbit gives rise to the secular changes which are on the order of $\epssa{1} (a_2/\qper)^2 [m_3 m_4/(m_3+m_4)^2]$, where $\epssa{1}$ is the magnitude of the quadrupole-order change (see \eq~\ref{eq:epssa}), and $a_2$ and $(m_3,m_4)$ are the companion binary orbital semimajor axis and component masses, respectively. Here, we largely ignored the fact that the inner orbits change dynamically during the encounter, i.e., we restricted to expressions to first order in the perturbation parameter $\epssa{i}$ except for the quadrupole order (see also \citealt{2019MNRAS.487.5630H}). Nevertheless, the analytic expressions generally agree with numerical integrations (\S~\ref{sect:num:bin}). 

\medskip \noindent 4. Most importantly, as shown by our analytic and numerical results, the `binarity' of orbit 2, when considering orbit 1, typically leads to only very small eccentricity and inclination changes. To good approximation, one can obtain the secular changes by using the analytic results for binary-single interactions \citep{1996MNRAS.282.1064H,2009ApJ...697..458S,2018MNRAS.476.4139H,2019ApJ...872..165G,2019MNRAS.487.5630H,2019MNRAS.488.5192H} and replacing the mass of the intruding unbound third body with the total mass of binary 2. In other words, the point-mass approximation works well in this case.

Several \textsc{Python} scripts implementing the two numerical integration methods and the analytical results as well as routines used to make all the plots in this paper, are freely available at the link given in \S~\ref{sect:num}.

\section*{Acknowledgements}
We thank the referee, Douglas Heggie, for a very helpful report. J.S. acknowledges support from the European Unions Horizon 2020 research and innovation programme under the Marie Sklodowska-Curie grant agreement No. 844629. Simulations in this paper made use of the \textsc{REBOUND} code which is freely available at \href{http://github.com/hannorein/rebound}{http://github.com/hannorein/rebound}. 

\bibliographystyle{mnras}
\bibliography{literature}

\begin{thebibliography}{}
\makeatletter
\relax
\def\mn@urlcharsother{\let\do\@makeother \do\$\do\&\do\#\do\^\do\_\do\%\do\~}
\def\mn@doi{\begingroup\mn@urlcharsother \@ifnextchar [ {\mn@doi@}
  {\mn@doi@[]}}
\def\mn@doi@[#1]#2{\def\@tempa{#1}\ifx\@tempa\@empty \href
  {http://dx.doi.org/#2} {doi:#2}\else \href {http://dx.doi.org/#2} {#1}\fi
  \endgroup}
\def\mn@eprint#1#2{\mn@eprint@#1:#2::\@nil}
\def\mn@eprint@arXiv#1{\href {http://arxiv.org/abs/#1} {{\tt arXiv:#1}}}
\def\mn@eprint@dblp#1{\href {http://dblp.uni-trier.de/rec/bibtex/#1.xml}
  {dblp:#1}}
\def\mn@eprint@#1:#2:#3:#4\@nil{\def\@tempa {#1}\def\@tempb {#2}\def\@tempc
  {#3}\ifx \@tempc \@empty \let \@tempc \@tempb \let \@tempb \@tempa \fi \ifx
  \@tempb \@empty \def\@tempb {arXiv}\fi \@ifundefined
  {mn@eprint@\@tempb}{\@tempb:\@tempc}{\expandafter \expandafter \csname
  mn@eprint@\@tempb\endcsname \expandafter{\@tempc}}}

\bibitem[\protect\citeauthoryear{{Abbott} et~al.,}{{Abbott}
  et~al.}{2016a}]{2016PhRvL.116f1102A}
{Abbott} B.~P.,  et~al., 2016a, \mn@doi [Physical Review Letters]
  {10.1103/PhysRevLett.116.061102}, \href
  {http://adsabs.harvard.edu/abs/2016PhRvL.116f1102A} {116, 061102}

\bibitem[\protect\citeauthoryear{{Abbott} et~al.,}{{Abbott}
  et~al.}{2016b}]{2016PhRvL.116x1103A}
{Abbott} B.~P.,  et~al., 2016b, \mn@doi [Physical Review Letters]
  {10.1103/PhysRevLett.116.241103}, \href
  {http://adsabs.harvard.edu/abs/2016PhRvL.116x1103A} {116, 241103}

\bibitem[\protect\citeauthoryear{{Abbott} et~al.,}{{Abbott}
  et~al.}{2017a}]{2017PhRvL.118v1101A}
{Abbott} B.~P.,  et~al., 2017a, \mn@doi [Physical Review Letters]
  {10.1103/PhysRevLett.118.221101}, \href
  {http://adsabs.harvard.edu/abs/2017PhRvL.118v1101A} {118, 221101}

\bibitem[\protect\citeauthoryear{{Abbott} et~al.,}{{Abbott}
  et~al.}{2017b}]{2017PhRvL.119n1101A}
{Abbott} B.~P.,  et~al., 2017b, \mn@doi [Physical Review Letters]
  {10.1103/PhysRevLett.119.141101}, \href
  {http://adsabs.harvard.edu/abs/2017PhRvL.119n1101A} {119, 141101}

\bibitem[\protect\citeauthoryear{{Abbott} et~al.,}{{Abbott}
  et~al.}{2017c}]{2017ApJ...848L..12A}
{Abbott} B.~P.,  et~al., 2017c, \mn@doi [\apjl] {10.3847/2041-8213/aa91c9},
  \href {http://adsabs.harvard.edu/abs/2017ApJ...848L..12A} {848, L12}

\bibitem[\protect\citeauthoryear{{Abbott} et~al.,}{{Abbott}
  et~al.}{2017d}]{2017ApJ...851L..35A}
{Abbott} B.~P.,  et~al., 2017d, \mn@doi [\apjl] {10.3847/2041-8213/aa9f0c},
  \href {http://adsabs.harvard.edu/abs/2017ApJ...851L..35A} {851, L35}

\bibitem[\protect\citeauthoryear{{Alexander}}{{Alexander}}{1986}]{1986JCoPh..64..195A}
{Alexander} M.~E.,  1986, \mn@doi [Journal of Computational Physics]
  {10.1016/0021-9991(86)90025-2}, \href
  {https://ui.adsabs.harvard.edu/abs/1986JCoPh..64..195A} {64, 195}

\bibitem[\protect\citeauthoryear{{Antognini} \& {Thompson}}{{Antognini} \&
  {Thompson}}{2016}]{2016MNRAS.456.4219A}
{Antognini} J.~M.~O.,  {Thompson} T.~A.,  2016, \mn@doi [\mnras]
  {10.1093/mnras/stv2938}, \href
  {http://adsabs.harvard.edu/abs/2016MNRAS.456.4219A} {456, 4219}

\bibitem[\protect\citeauthoryear{{Bacon}, {Sigurdsson}  \& {Davies}}{{Bacon}
  et~al.}{1996}]{1996MNRAS.281..830B}
{Bacon} D.,  {Sigurdsson} S.,   {Davies} M.~B.,  1996, \mn@doi [\mnras]
  {10.1093/mnras/281.3.830}, \href
  {https://ui.adsabs.harvard.edu/abs/1996MNRAS.281..830B} {281, 830}

\bibitem[\protect\citeauthoryear{{Davies}, {Benz}  \& {Hills}}{{Davies}
  et~al.}{1993}]{1993ApJ...411..285D}
{Davies} M.~B.,  {Benz} W.,   {Hills} J.~G.,  1993, \mn@doi [\apj]
  {10.1086/172828}, \href
  {https://ui.adsabs.harvard.edu/abs/1993ApJ...411..285D} {411, 285}

\bibitem[\protect\citeauthoryear{{Eggleton}}{{Eggleton}}{2006}]{2006epbm.book.....E}
{Eggleton} P.,  2006, {Evolutionary Processes in Binary and Multiple Stars}

\bibitem[\protect\citeauthoryear{{Geller}, {Leigh}, {Giersz}, {Kremer}  \&
  {Rasio}}{{Geller} et~al.}{2019}]{2019ApJ...872..165G}
{Geller} A.~M.,  {Leigh} N. W.~C.,  {Giersz} M.,  {Kremer} K.,   {Rasio} F.~A.,
   2019, \mn@doi [\apj] {10.3847/1538-4357/ab0214}, \href
  {https://ui.adsabs.harvard.edu/\#abs/2019ApJ...872..165G} {872, 165}

\bibitem[\protect\citeauthoryear{{Goodman} \& {Hut}}{{Goodman} \&
  {Hut}}{1993}]{1993ApJ...403..271G}
{Goodman} J.,  {Hut} P.,  1993, \mn@doi [\apj] {10.1086/172200}, \href
  {https://ui.adsabs.harvard.edu/abs/1993ApJ...403..271G} {403, 271}

\bibitem[\protect\citeauthoryear{{Hamers}}{{Hamers}}{2018}]{2018MNRAS.476.4139H}
{Hamers} A.~S.,  2018, \mn@doi [\mnras] {10.1093/mnras/sty428}, \href
  {https://ui.adsabs.harvard.edu/\#abs/2018MNRAS.476.4139H} {476, 4139}

\bibitem[\protect\citeauthoryear{{Hamers} \& {Portegies Zwart}}{{Hamers} \&
  {Portegies Zwart}}{2016}]{2016MNRAS.459.2827H}
{Hamers} A.~S.,  {Portegies Zwart} S.~F.,  2016, \mn@doi [\mnras]
  {10.1093/mnras/stw784}, \href
  {http://adsabs.harvard.edu/abs/2016MNRAS.459.2827H} {459, 2827}

\bibitem[\protect\citeauthoryear{{Hamers} \& {Samsing}}{{Hamers} \&
  {Samsing}}{2019a}]{2019MNRAS.487.5630H}
{Hamers} A.~S.,  {Samsing} J.,  2019a, \mn@doi [\mnras]
  {10.1093/mnras/stz1646}, \href
  {https://ui.adsabs.harvard.edu/abs/2019MNRAS.487.5630H} {487, 5630}

\bibitem[\protect\citeauthoryear{{Hamers} \& {Samsing}}{{Hamers} \&
  {Samsing}}{2019b}]{2019MNRAS.488.5192H}
{Hamers} A.~S.,  {Samsing} J.,  2019b, \mn@doi [\mnras]
  {10.1093/mnras/stz2029}, \href
  {https://ui.adsabs.harvard.edu/abs/2019MNRAS.488.5192H} {488, 5192}

\bibitem[\protect\citeauthoryear{{Hamers}, {Perets}, {Antonini}  \& {Portegies
  Zwart}}{{Hamers} et~al.}{2015}]{2015MNRAS.449.4221H}
{Hamers} A.~S.,  {Perets} H.~B.,  {Antonini} F.,   {Portegies Zwart} S.~F.,
  2015, \mn@doi [\mnras] {10.1093/mnras/stv452}, \href
  {https://ui.adsabs.harvard.edu/abs/2015MNRAS.449.4221H} {449, 4221}

\bibitem[\protect\citeauthoryear{{Heggie}}{{Heggie}}{1975}]{1975MNRAS.173..729H}
{Heggie} D.~C.,  1975, \mn@doi [\mnras] {10.1093/mnras/173.3.729}, \href
  {http://adsabs.harvard.edu/abs/1975MNRAS.173..729H} {173, 729}

\bibitem[\protect\citeauthoryear{{Heggie} \& {Hut}}{{Heggie} \&
  {Hut}}{1993}]{1993ApJS...85..347H}
{Heggie} D.~C.,  {Hut} P.,  1993, \mn@doi [\apjs] {10.1086/191768}, \href
  {https://ui.adsabs.harvard.edu/abs/1993ApJS...85..347H} {85, 347}

\bibitem[\protect\citeauthoryear{{Heggie} \& {Rasio}}{{Heggie} \&
  {Rasio}}{1996}]{1996MNRAS.282.1064H}
{Heggie} D.~C.,  {Rasio} F.~A.,  1996, \mn@doi [\mnras]
  {10.1093/mnras/282.3.1064}, \href
  {http://adsabs.harvard.edu/abs/1996MNRAS.282.1064H} {282, 1064}

\bibitem[\protect\citeauthoryear{{Heggie}, {Hut}  \& {McMillan}}{{Heggie}
  et~al.}{1996}]{1996ApJ...467..359H}
{Heggie} D.~C.,  {Hut} P.,   {McMillan} S.~L.~W.,  1996, \mn@doi [\apj]
  {10.1086/177611}, \href {http://adsabs.harvard.edu/abs/1996ApJ...467..359H}
  {467, 359}

\bibitem[\protect\citeauthoryear{{Hoffer}}{{Hoffer}}{1983}]{1983AJ.....88.1420H}
{Hoffer} J.~B.,  1983, \mn@doi [\aj] {10.1086/113431}, \href
  {https://ui.adsabs.harvard.edu/abs/1983AJ.....88.1420H} {88, 1420}

\bibitem[\protect\citeauthoryear{{Hut}}{{Hut}}{1983}]{1983ApJ...268..342H}
{Hut} P.,  1983, \mn@doi [\apj] {10.1086/160957}, \href
  {http://adsabs.harvard.edu/abs/1983ApJ...268..342H} {268, 342}

\bibitem[\protect\citeauthoryear{{Hut}}{{Hut}}{1993}]{1993ApJ...403..256H}
{Hut} P.,  1993, \mn@doi [\apj] {10.1086/172199}, \href
  {http://adsabs.harvard.edu/abs/1993ApJ...403..256H} {403, 256}

\bibitem[\protect\citeauthoryear{{Hut} \& {Bahcall}}{{Hut} \&
  {Bahcall}}{1983}]{1983ApJ...268..319H}
{Hut} P.,  {Bahcall} J.~N.,  1983, \mn@doi [\apj] {10.1086/160956}, \href
  {http://adsabs.harvard.edu/abs/1983ApJ...268..319H} {268, 319}

\bibitem[\protect\citeauthoryear{{Hut}, {McMillan}  \& {Romani}}{{Hut}
  et~al.}{1992}]{1992ApJ...389..527H}
{Hut} P.,  {McMillan} S.,   {Romani} R.~W.,  1992, \mn@doi [\apj]
  {10.1086/171229}, \href
  {https://ui.adsabs.harvard.edu/abs/1992ApJ...389..527H} {389, 527}

\bibitem[\protect\citeauthoryear{{Kimpson}, {Spera}, {Mapelli}  \&
  {Ziosi}}{{Kimpson} et~al.}{2016}]{2016MNRAS.463.2443K}
{Kimpson} T.~O.,  {Spera} M.,  {Mapelli} M.,   {Ziosi} B.~M.,  2016, \mn@doi
  [\mnras] {10.1093/mnras/stw2085}, \href
  {http://adsabs.harvard.edu/abs/2016MNRAS.463.2443K} {463, 2443}

\bibitem[\protect\citeauthoryear{{Kocsis} \& {Levin}}{{Kocsis} \&
  {Levin}}{2012}]{2012PhRvD..85l3005K}
{Kocsis} B.,  {Levin} J.,  2012, \mn@doi [\prd] {10.1103/PhysRevD.85.123005},
  \href {https://ui.adsabs.harvard.edu/abs/2012PhRvD..85l3005K} {85, 123005}

\bibitem[\protect\citeauthoryear{{Lei}}{{Lei}}{2019}]{2019MNRAS.490.4756L}
{Lei} H.,  2019, \mn@doi [\mnras] {10.1093/mnras/stz2917}, \href
  {https://ui.adsabs.harvard.edu/abs/2019MNRAS.490.4756L} {490, 4756}

\bibitem[\protect\citeauthoryear{{Leigh} \& {Geller}}{{Leigh} \&
  {Geller}}{2012}]{2012MNRAS.425.2369L}
{Leigh} N.,  {Geller} A.~M.,  2012, \mn@doi [\mnras]
  {10.1111/j.1365-2966.2012.21689.x}, \href
  {http://adsabs.harvard.edu/abs/2012MNRAS.425.2369L} {425, 2369}

\bibitem[\protect\citeauthoryear{{Leigh} \& {Geller}}{{Leigh} \&
  {Geller}}{2015}]{2015MNRAS.450.1724L}
{Leigh} N.~W.~C.,  {Geller} A.~M.,  2015, \mn@doi [\mnras]
  {10.1093/mnras/stv685}, \href
  {http://adsabs.harvard.edu/abs/2015MNRAS.450.1724L} {450, 1724}

\bibitem[\protect\citeauthoryear{{Leigh} \& {Sills}}{{Leigh} \&
  {Sills}}{2011}]{2011MNRAS.410.2370L}
{Leigh} N.,  {Sills} A.,  2011, \mn@doi [\mnras]
  {10.1111/j.1365-2966.2010.17609.x}, \href
  {https://ui.adsabs.harvard.edu/abs/2011MNRAS.410.2370L} {410, 2370}

\bibitem[\protect\citeauthoryear{{Leigh}, {Geller}, {Shara}, {Garland},
  {Clees-Baron}  \& {Ahmed}}{{Leigh} et~al.}{2017}]{2017MNRAS.471.1830L}
{Leigh} N. W.~C.,  {Geller} A.~M.,  {Shara} M.~M.,  {Garland} J.,
  {Clees-Baron} H.,   {Ahmed} A.,  2017, \mn@doi [\mnras]
  {10.1093/mnras/stx1704}, \href
  {https://ui.adsabs.harvard.edu/abs/2017MNRAS.471.1830L} {471, 1830}

\bibitem[\protect\citeauthoryear{{Leigh}, {Geller}, {Shara}, {Baugher},
  {Hierro}, {Ferreira}  \& {Teperino}}{{Leigh}
  et~al.}{2018}]{2018MNRAS.480.3062L}
{Leigh} N. W.~C.,  {Geller} A.~M.,  {Shara} M.~M.,  {Baugher} L.,  {Hierro} V.,
   {Ferreira} D.,   {Teperino} E.,  2018, \mn@doi [\mnras]
  {10.1093/mnras/sty2046}, \href
  {https://ui.adsabs.harvard.edu/abs/2018MNRAS.480.3062L} {480, 3062}

\bibitem[\protect\citeauthoryear{{Leonard}}{{Leonard}}{1989}]{1989AJ.....98..217L}
{Leonard} P. J.~T.,  1989, \mn@doi [\aj] {10.1086/115138}, \href
  {https://ui.adsabs.harvard.edu/abs/1989AJ.....98..217L} {98, 217}

\bibitem[\protect\citeauthoryear{{Li} \& {Adams}}{{Li} \&
  {Adams}}{2015}]{2015MNRAS.448..344L}
{Li} G.,  {Adams} F.~C.,  2015, \mn@doi [\mnras] {10.1093/mnras/stv012}, \href
  {http://adsabs.harvard.edu/abs/2015MNRAS.448..344L} {448, 344}

\bibitem[\protect\citeauthoryear{{Mapelli}}{{Mapelli}}{2016}]{2016MNRAS.459.3432M}
{Mapelli} M.,  2016, \mn@doi [\mnras] {10.1093/mnras/stw869}, \href
  {http://adsabs.harvard.edu/abs/2016MNRAS.459.3432M} {459, 3432}

\bibitem[\protect\citeauthoryear{{McMillan} \& {Hut}}{{McMillan} \&
  {Hut}}{1994}]{1994ApJ...427..793M}
{McMillan} S.,  {Hut} P.,  1994, \mn@doi [\apj] {10.1086/174186}, \href
  {https://ui.adsabs.harvard.edu/abs/1994ApJ...427..793M} {427, 793}

\bibitem[\protect\citeauthoryear{{McMillan} \& {Hut}}{{McMillan} \&
  {Hut}}{1996}]{1996ApJ...467..348M}
{McMillan} S. L.~W.,  {Hut} P.,  1996, \mn@doi [\apj] {10.1086/177610}, \href
  {https://ui.adsabs.harvard.edu/abs/1996ApJ...467..348M} {467, 348}

\bibitem[\protect\citeauthoryear{{Mikkola}}{{Mikkola}}{1983}]{1983MNRAS.203.1107M}
{Mikkola} S.,  1983, \mn@doi [\mnras] {10.1093/mnras/203.4.1107}, \href
  {https://ui.adsabs.harvard.edu/abs/1983MNRAS.203.1107M} {203, 1107}

\bibitem[\protect\citeauthoryear{{Mikkola}}{{Mikkola}}{1984a}]{1984MNRAS.207..115M}
{Mikkola} S.,  1984a, \mn@doi [\mnras] {10.1093/mnras/207.1.115}, \href
  {https://ui.adsabs.harvard.edu/abs/1984MNRAS.207..115M} {207, 115}

\bibitem[\protect\citeauthoryear{{Mikkola}}{{Mikkola}}{1984b}]{1984MNRAS.208...75M}
{Mikkola} S.,  1984b, \mn@doi [\mnras] {10.1093/mnras/208.1.75}, \href
  {https://ui.adsabs.harvard.edu/abs/1984MNRAS.208...75M} {208, 75}

\bibitem[\protect\citeauthoryear{{O'Leary}, {Rasio}, {Fregeau}, {Ivanova}  \&
  {O'Shaughnessy}}{{O'Leary} et~al.}{2006}]{2006ApJ...637..937O}
{O'Leary} R.~M.,  {Rasio} F.~A.,  {Fregeau} J.~M.,  {Ivanova} N.,
  {O'Shaughnessy} R.,  2006, \mn@doi [\apj] {10.1086/498446}, \href
  {http://adsabs.harvard.edu/abs/2006ApJ...637..937O} {637, 937}

\bibitem[\protect\citeauthoryear{{Portegies Zwart} \& {McMillan}}{{Portegies
  Zwart} \& {McMillan}}{2000}]{2000ApJ...528L..17P}
{Portegies Zwart} S.~F.,  {McMillan} S.~L.~W.,  2000, \mn@doi [\apjl]
  {10.1086/312422}, \href {http://adsabs.harvard.edu/abs/2000ApJ...528L..17P}
  {528, L17}

\bibitem[\protect\citeauthoryear{{Rasio}, {McMillan}  \& {Hut}}{{Rasio}
  et~al.}{1995}]{1995ApJ...438L..33R}
{Rasio} F.~A.,  {McMillan} S.,   {Hut} P.,  1995, \mn@doi [\apjl]
  {10.1086/187708}, \href
  {https://ui.adsabs.harvard.edu/abs/1995ApJ...438L..33R} {438, L33}

\bibitem[\protect\citeauthoryear{{Rein} \& {Liu}}{{Rein} \&
  {Liu}}{2012}]{2012A&A...537A.128R}
{Rein} H.,  {Liu} S.~F.,  2012, \mn@doi [\aap] {10.1051/0004-6361/201118085},
  \href {https://ui.adsabs.harvard.edu/abs/2012A&A...537A.128R} {537, A128}

\bibitem[\protect\citeauthoryear{{Rein} \& {Spiegel}}{{Rein} \&
  {Spiegel}}{2015}]{2015MNRAS.446.1424R}
{Rein} H.,  {Spiegel} D.~S.,  2015, \mn@doi [\mnras] {10.1093/mnras/stu2164},
  \href {https://ui.adsabs.harvard.edu/abs/2015MNRAS.446.1424R} {446, 1424}

\bibitem[\protect\citeauthoryear{{Rodriguez}, {Morscher}, {Pattabiraman},
  {Chatterjee}, {Haster}  \& {Rasio}}{{Rodriguez}
  et~al.}{2015}]{2015PhRvL.115e1101R}
{Rodriguez} C.~L.,  {Morscher} M.,  {Pattabiraman} B.,  {Chatterjee} S.,
  {Haster} C.-J.,   {Rasio} F.~A.,  2015, \mn@doi [Physical Review Letters]
  {10.1103/PhysRevLett.115.051101}, \href
  {http://adsabs.harvard.edu/abs/2015PhRvL.115e1101R} {115, 051101}

\bibitem[\protect\citeauthoryear{{Rodriguez}, {Chatterjee}  \&
  {Rasio}}{{Rodriguez} et~al.}{2016}]{2016PhRvD..93h4029R}
{Rodriguez} C.~L.,  {Chatterjee} S.,   {Rasio} F.~A.,  2016, \mn@doi [\prd]
  {10.1103/PhysRevD.93.084029}, \href
  {http://adsabs.harvard.edu/abs/2016PhRvD..93h4029R} {93, 084029}

\bibitem[\protect\citeauthoryear{{Rodriguez}, {Amaro-Seoane}, {Chatterjee}  \&
  {Rasio}}{{Rodriguez} et~al.}{2018}]{2018PhRvL.120o1101R}
{Rodriguez} C.~L.,  {Amaro-Seoane} P.,  {Chatterjee} S.,   {Rasio} F.~A.,
  2018, \mn@doi [Physical Review Letters] {10.1103/PhysRevLett.120.151101},
  \href {http://adsabs.harvard.edu/abs/2018PhRvL.120o1101R} {120, 151101}

\bibitem[\protect\citeauthoryear{{Ryu}, {Leigh}  \& {Perna}}{{Ryu}
  et~al.}{2017}]{2017MNRAS.467.4447R}
{Ryu} T.,  {Leigh} N. W.~C.,   {Perna} R.,  2017, \mn@doi [\mnras]
  {10.1093/mnras/stx395}, \href
  {https://ui.adsabs.harvard.edu/abs/2017MNRAS.467.4447R} {467, 4447}

\bibitem[\protect\citeauthoryear{{Samsing}}{{Samsing}}{2018}]{2018PhRvD..97j3014S}
{Samsing} J.,  2018, \mn@doi [\prd] {10.1103/PhysRevD.97.103014}, \href
  {http://adsabs.harvard.edu/abs/2018PhRvD..97j3014S} {97, 103014}

\bibitem[\protect\citeauthoryear{{Samsing} \& {Ramirez-Ruiz}}{{Samsing} \&
  {Ramirez-Ruiz}}{2017}]{2017ApJ...840L..14S}
{Samsing} J.,  {Ramirez-Ruiz} E.,  2017, \mn@doi [\apjl]
  {10.3847/2041-8213/aa6f0b}, \href
  {http://adsabs.harvard.edu/abs/2017ApJ...840L..14S} {840, L14}

\bibitem[\protect\citeauthoryear{{Samsing}, {D'Orazio}, {Askar}  \&
  {Giersz}}{{Samsing} et~al.}{2018a}]{2018arXiv180208654S}
{Samsing} J.,  {D'Orazio} D.~J.,  {Askar} A.,   {Giersz} M.,  2018a, arXiv
  e-prints, \href {https://ui.adsabs.harvard.edu/abs/2018arXiv180208654S} {p.
  arXiv:1802.08654}

\bibitem[\protect\citeauthoryear{{Samsing}, {MacLeod}  \&
  {Ramirez-Ruiz}}{{Samsing} et~al.}{2018b}]{2018ApJ...853..140S}
{Samsing} J.,  {MacLeod} M.,   {Ramirez-Ruiz} E.,  2018b, \mn@doi [\apj]
  {10.3847/1538-4357/aaa715}, \href
  {https://ui.adsabs.harvard.edu/abs/2018ApJ...853..140S} {853, 140}

\bibitem[\protect\citeauthoryear{{Samsing}, {Askar}  \& {Giersz}}{{Samsing}
  et~al.}{2018c}]{2018ApJ...855..124S}
{Samsing} J.,  {Askar} A.,   {Giersz} M.,  2018c, \mn@doi [\apj]
  {10.3847/1538-4357/aaab52}, \href
  {https://ui.adsabs.harvard.edu/abs/2018ApJ...855..124S} {855, 124}

\bibitem[\protect\citeauthoryear{{Samsing}, {Hamers}  \& {Tyles}}{{Samsing}
  et~al.}{2019}]{2019PhRvD.100d3010S}
{Samsing} J.,  {Hamers} A.~S.,   {Tyles} J.~G.,  2019, \mn@doi [\prd]
  {10.1103/PhysRevD.100.043010}, \href
  {https://ui.adsabs.harvard.edu/abs/2019PhRvD.100d3010S} {100, 043010}

\bibitem[\protect\citeauthoryear{{Sigurdsson} \& {Hernquist}}{{Sigurdsson} \&
  {Hernquist}}{1993}]{1993Natur.364..423S}
{Sigurdsson} S.,  {Hernquist} L.,  1993, \mn@doi [\nat] {10.1038/364423a0},
  \href {http://adsabs.harvard.edu/abs/1993Natur.364..423S} {364, 423}

\bibitem[\protect\citeauthoryear{{Sigurdsson} \& {Phinney}}{{Sigurdsson} \&
  {Phinney}}{1993}]{1993ApJ...415..631S}
{Sigurdsson} S.,  {Phinney} E.~S.,  1993, \mn@doi [\apj] {10.1086/173190},
  \href {https://ui.adsabs.harvard.edu/abs/1993ApJ...415..631S} {415, 631}

\bibitem[\protect\citeauthoryear{{Spurzem}, {Giersz}, {Heggie}  \&
  {Lin}}{{Spurzem} et~al.}{2009}]{2009ApJ...697..458S}
{Spurzem} R.,  {Giersz} M.,  {Heggie} D.~C.,   {Lin} D.~N.~C.,  2009, \mn@doi
  [\apj] {10.1088/0004-637X/697/1/458}, \href
  {http://adsabs.harvard.edu/abs/2009ApJ...697..458S} {697, 458}

\bibitem[\protect\citeauthoryear{{Zevin}, {Samsing}, {Rodriguez}, {Haster}  \&
  {Ramirez-Ruiz}}{{Zevin} et~al.}{2019}]{2019ApJ...871...91Z}
{Zevin} M.,  {Samsing} J.,  {Rodriguez} C.,  {Haster} C.-J.,   {Ramirez-Ruiz}
  E.,  2019, \mn@doi [\apj] {10.3847/1538-4357/aaf6ec}, \href
  {https://ui.adsabs.harvard.edu/abs/2019ApJ...871...91Z} {871, 91}

\bibitem[\protect\citeauthoryear{{Ziosi}, {Mapelli}, {Branchesi}  \&
  {Tormen}}{{Ziosi} et~al.}{2014}]{2014MNRAS.441.3703Z}
{Ziosi} B.~M.,  {Mapelli} M.,  {Branchesi} M.,   {Tormen} G.,  2014, \mn@doi
  [\mnras] {10.1093/mnras/stu824}, \href
  {http://adsabs.harvard.edu/abs/2014MNRAS.441.3703Z} {441, 3703}

\makeatother
\end{thebibliography}

\label{lastpage}

\end{document}